\documentclass[preprint,prb,superscriptaddress,aps]{revtex4}
\usepackage{graphicx}
\usepackage{float}
\usepackage{dcolumn}
\usepackage{amsmath,amssymb}
\usepackage{multirow}
\usepackage{boxedminipage}
\usepackage{array}
\usepackage{flafter}
\usepackage[mathscr]{euscript}
\usepackage[tight,TABTOPCAP]{subfigure}
\usepackage[normalsize]{caption}
\usepackage{verbatim}
\usepackage{xcolor}
\usepackage[normalem]{ulem}
\newcolumntype{.}{D{.}{.}{1}}
\begin{document}

\title{
A kinetic model to simulate charge flow through an electrochemical half cell
}

\author{Diego Veloza-Diaz}
\affiliation{Institut f{\"u}r Physik, Johannes Gutenberg-Universit{\"a}t Mainz, Staudingerweg 9, 55128 Mainz, Germany}
\author{Friederike Schmid}
\affiliation{Institut f{\"u}r Physik, Johannes Gutenberg-Universit{\"a}t Mainz, Staudingerweg 9, 55128 Mainz, Germany}
\author{Robinson Cortes-Huerto$^{**}$}
\affiliation{Max Planck Institute for Polymer Research, Ackermannweg 10, 55128, Mainz, Germany}
\author{Pietro Ballone}
\affiliation{Max Planck Institute for Polymer Research, Ackermannweg 10, 55128, Mainz, Germany}
\author{Nancy C. Forero-Martinez$^{*}$}
\affiliation{Institut f{\"u}r Physik, Johannes Gutenberg-Universit{\"a}t Mainz, Staudingerweg 9, 55128 Mainz, Germany}

\begin{abstract}
A kinetic model of the electron transfer at the electrode / electrolyte solution interface is developed, implemented in a Monte 
Carlo framework, and applied to simulate this process in idealised systems consisting of the primitive model of electrolyte 
solutions limited by an impenetrable conducting surface. In the present implementation, a charged, spherical interface 
surrounding an equally spherical sample of electrolyte solution is introduced to model a single-electrode system, providing the 
computational analog to the conceptual half-cell picture that is widely used in electrochemistry. The electron transfer itself 
is described as a simple surface hopping process underlying a first order reaction corresponding to one of the coupled M/M$^+$ 
and X$^-$/X half reactions. Then, the electron transfer at the interface is combined with the self-diffusion of ions in the 
electrolyte solutions whose role is to supply reagents and disperse products, allowing the system to settle in a stationary 
non-equilibrium state. Simulations for the primitive model of electrolyte in contact with a charged impenetrable surface show 
that, after a brief transient, the samples sustain a steady current through the electrolyte solution. The results quantify the 
dependence of the current on: the overall charge of the electrode, the electrolyte concentration, the solvent viscosity and the 
kinetic parameter $k_e$ that represents the rate of the electron transfer for each ion in contact with the electrode. Since the 
simulated interface is very idealised, strategies to overcome the limitations of the present model are outlined and briefly 
discussed. 
\vskip 1.0truecm
\noindent
{\bf $^{ *}$ Corresponding author: nforerom@uni-mainz.de}\\
{\bf $^{**}$ Corresponding author: corteshu@mpip-mainz.mpg.de}
\end{abstract}

\maketitle

\section{Introduction}
\label{intro}
The transfer of electrons across the interface between a metal electrode and an electrolyte solution is a  complex and
fascinating phenomenon of great conceptual and practical interest, which underlies the functioning of both batteries and
electrochemical cells, and plays a role in a variety of other topics, from catalysis to corrosion.\cite{inter} For this reason, 
a sizeable portion of electrochemistry is devoted to this phenomenon,\cite{bockris} whose full understanding, however, is still 
elusive. 

The feeling of incompleteness concerns especially the theoretical and computational description, since the electron transfer at 
the electrode / electrolyte interface encompasses a broad range of aspects, whose analysis requires approaches that are 
difficult to combine into a single scheme. Exemplary in this respect are the joint quantum / classical mechanics aspect of the 
electron 
transfer between electrode and ions in solution, the interplay of statistical mechanics and electronic structure methods that 
are both essential for a full description of the interface and of its electrification, and the important role of steady state 
non-equilibrium conditions that characterise the normal operation of every electrochemical  device. 

\begin{figure}[!htb]
\begin{minipage}[c]{\textwidth}
\vskip 0.7truecm
\begin{center}
\includegraphics[scale=0.70,angle=-0]{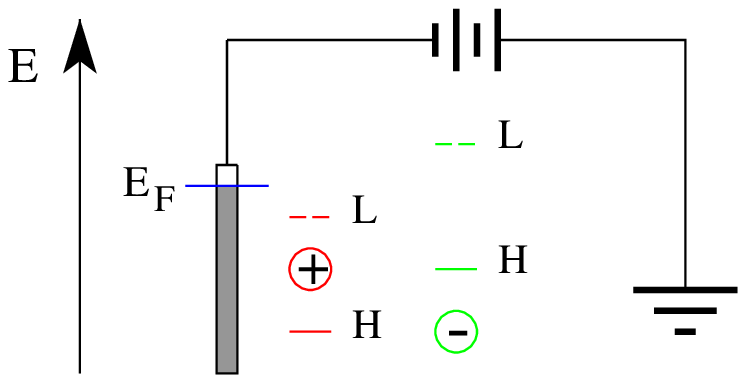}
\vskip 2.0truecm
\caption{Schematic diagram of the half-cell model. The electron energy is measured on the common scale indicated by the vertical
axis on the left. The atomic-like energy levels on the ions are indicated by a continuous (HOMO: highest occupied molecular 
orbital) and dashed (LUMO: lowest unoccupied molecular orbital) segment, respectively. 
}
\label{scheme}
\end{center}
\end{minipage}
\end{figure}

For the sake of definiteness, in what follows the attention will be focused on the electrochemical cell case, in which the flow 
of electrons forced, for instance, by a battery drives a chemical reaction. The model discussed in the following sections, in 
particular, aims at reproducing the behaviour of one half of the device, in which electrons exiting the cathode reduce (M$^+$ + 
e $\rightarrow$ M) the cations in solution. This mode of operation assumes that the energy of the electronic states on the metal
electrode and ions in solution reflects the distribution schematically given in Fig.~\ref{scheme}. Extending the model to the 
oxidation taking place at the anion is trivial, and the two half-reactions can be combined to model either an electrochemical
or a Voltaic cell.

A crucial portion of the whole problem concerns the inner interfacial region (which could be identified with the Helmholtz 
layer\cite{atk}), encompassing a thin layer on both sides of the geometric interface. In this region, ions change their 
oxidation state, while shedding or greatly reorganising their hydration shell, a process associated with a free energy barrier 
which is known as {\it reorganisation energy}.\cite{marcus, marcusb, hush} This has to take place while the transferred electron
adapts to the qualitatively different condition on the two sides of the interface, consisting of localised orbitals and sharp 
energy eigenvalues on the electrolyte side, versus delocalised orbitals and energy bands on the metal side. The combination of 
these effects, i.e., the reorganisation energy and the electronic transition, determine the electron transfer rate $k_e$ whose 
computation is possibly the most challenging part of the entire problem.\cite{schmic} Also important, however, is to understand 
how the electronic processes at the interface couple to the current that has to flow in the electrolyte solution to keep the 
system in a steady state. Crucial, in this respect, is understanding the influence of the interface electrification, represented
primarily by the interfacial double layer, on the electron transfer process and on its coupling with the current flowing through
the electrolyte. Covering all these different aspects presents the additional challenge of matching phenomena developing over 
different size and time scales.

Ideally, one would like to have a unique framework suitable to follow the time evolution of an electrochemical cell, joining all
the different aspects that it encompasses. In perspective, there is the feeling that ab-initio approaches, and ab-initio 
simulation methods in particular,\cite{price} hold the promise to fulfill this role. However, both the cost  and the conceptual 
limitations of current methods that involve density functional theory (DFT) in its computationally viable approximations, 
prevent ab-initio approaches to fully meet these expectations. First of all, most ab-initio / molecular dynamics simulation 
methods require that the time evolution of the system is adiabatic, i.e., it takes place on a single Born-Oppenheimer potential
energy surface. Hence, popular ab-initio simulation methods are devised in such a way to describe the Newtonian dynamics of 
atoms and ions, while electrons follow a fake dynamics whose aim is to keep them at or close to the minimum of the functional 
which gives the potential energy of the electron subsystem as a function of a continuous 3D distribution function that 
represents the electron density. As a result, it is difficult to interpret the time evolution of the electron density as given 
by these methods in terms of a real-time dynamics, if not in a very smeared and averaged way. In particular, ab-initio 
simulation approaches lose the particle-like dynamics of the electrons, that include the granularity, randomness and jump-like 
features that underly the time evolution of electrons in electrochemical systems. Because of these reasons, it is still 
desirable to develop computationally less demanding methods, even at the cost of introducing drastic simplifications.

In undertaking this task, it might be useful to summarise the aspects that the model is expected to reproduce. First, the charge
transfer is a quantum mechanical process that discontinuously changes the oxidation state of the ions. The granularity of the 
process, arising from the multitude of discrete electron transfer events, is a crucial aspect that the model need to retain to
faithfully reproduce correlations among transfer events, as well as the noise inherent in the flow of charge through the system.
The electronic charge of the metal, in turn, is quickly restored by a generator of continuous electric current, such as a 
battery, and, to all purposes, it can be considered constant during the electrochemical device operation. The surface charge 
$\sigma$ of the electrode is not the primary cause of the electron transfer, which is instead driven by the relative energy and 
spatial superposition of electron states on the metal and on the ions, see Fig.~\ref{scheme}. The surface charge, however, is 
still an important parameter, as confirmed by the simulation results shown below. Moreover, its value as a function of the 
applied potential defines the potential of zero (surface) charge or PZC, for which $\sigma$ vanishes, which is a very important 
reference state for the electrode.\cite{pzc} 

The electronic part of the charge transfer concerns ions in the immediate vicinity of the electrode, a population that in the 
following will be labelled as {\it active}. In most cases the charge transfer has to overcome a free energy barrier. Hence, this
process is slow on the typical time scale of other microscopic processes taking place in the system. Defining the rates $k_e$ 
as the average number of events {\it per active ion} and per unit of time, the expected values are somewhat less than one event 
per active ion per nanosecond. Despite being slow, the charge transfer would eventually deplete the inner layer populated by 
ions in close contact with the electrode, unless another mechanism, i.e., self-diffusion, to some extent compensates the charge 
transfer, tending to regenerate the equilibrium distribution of ions in the system. The compensation, however, is not complete. 
At the conditions of the simulations presented below, the alteration of the charge density profile in the vicinity of the 
electrode due to electron transfer, although fairly small, is apparent and can be accurately computed. It is important to remark
that, at the same conditions, the ionic diffusion in the electrolyte solution runs on a faster time scale than the electron 
transfer process. Detailed atomistic simulations\cite{turq} show that the velocity autocorrelation function for simple 
electrolyte ions decays exponentially with a time constant of about $1$ ps. This is therefore the time needed to a diffusing ion
to lose memory of its direction, making its trajectory practically and conceptually similar to a Brownian random walk. It is 
important to remark that this (ps) time scale is orders of magnitude faster than the electron transfer time scale given by the 
inverse of $k_e$.

The crucial problem in developing the model consists in synchronising the two essential aspects, i.e., the charge 
transfer\cite{schmic} and the ionic diffusion and electrical conductivity on the electrolyte side,\cite{valleau} mixing them in 
the correct way that reflects the faster time scale of diffusion with respect to the kinetics of the single electron transfer. 
To this aim, we will resort to concepts inspired by the well known kinetic Monte Carlo algorithm,\cite{KMC, KMCb} but also 
sharing aspects, algorithms and limitations with standard methods to simulate chemical kinetics. Their combination aims at 
covering long time intervals, retaining sufficient temporal and spatial resolution to faithfully reproduce the mutual effects of
the two dominant processes.

As a side issue, a new geometric representation of electrode / electrolyte interfaces has been introduced, motivated by the 
following considerations. Electrochemical cells and batteries always involve two interfaces of opposite polarity. In real, 
macroscopic systems, the two sides of an electrochemical cell are both in equilibrium with a common neutral bulk electrolyte 
solution. This has motivated the usage in electrochemistry of the half cell (or half battery) concept. We attempt to develop a 
similar conceptual tool, introducing a spherical electrode which, combined with the grand canonical Monte Carlo (GCMC) 
simulation method is meant to represent a single one of the two sides of any electrochemical cell. As it is often the case, this
model has both advantages and a few disadvantages with respect to the planar slab model of the electrochemical cell, as will 
be briefly discussed in the paper. Nevertheless, the new electrode geometry is an additional element in the computational 
toolkit available for electrochemical studies, and in the present study, it allows us to focus on the narrow interfacial 
region, excluding any possible interference with the other side of the cell, despite the still microscopic/mesoscopic size of 
the simulation cell.

\section{Model and Method}
\label{method}
For the sake of clarity and simplicity, the computational approach is presented and tested for a very idealised model of
metal/electrolyte interface, similar to those used in early stages of computer simulation,\cite{torrie} but many of the details 
and realism of current atomistic force-field models can easily be reintroduced at a later stage.

The electrolyte-water solution is represented by the primitive model,\cite{moder} which is an implicit solvent model, consisting
of an assembly of $N$ charged hard spheres. In other terms, the water solvent enters the model only as the dielectric medium 
that decreases the Coulomb interaction among ions by a factor equal to the static dielectric constant $\epsilon_0=78.5$. The 
ion-ion interaction is of the hard-sphere type, supplemented by Coulomb charges at the centre of each sphere. In general, the 
choice of the hard sphere diameters, i.e., $d_{++}$, $d_{+-}$, $d_{--}$ for cation-cation, cation-anion and anion-anion pairs, 
respectively, is not restricted by symmetry ($d_{++}=d_{--}$) or additivity ($d_{+-}=(d_{++}+d_{--})/2$) conditions. The system 
Hamiltonian is:
\begin{equation}
\hat{H}= \sum_i \frac{\bf p_i^2}{2m_i}+U_N (\{ {\bf r}_i; i=1, ..., N\})
\label{hamilt0}
\end{equation}
In this equation, $\{ {\bf p_i}; i=1, ..., N\}$ and $\{ {\bf r_i}; i=1, ..., N\}$ represent the momenta and positions, 
respectively, of the $N$ ions, whose mass is $\{m_i;  i=1, ..., N\}$ and charge is $\{q_i;  i=1, ..., N\}$. The potential 
energy $U(\{ {\bf r_i}; i=1, ..., N \})$ is given by:
\begin{equation}
U_N (\{ {\bf r_i}; i=1, ..., N \})=\frac{1}{2}{\sum_{i\neq j=1}^N} v(\mid {\bf r_i-r_j}\mid)
\end{equation}
and:
\begin{equation}
v(\mid {\bf r_i-r_j}\mid)=\left\{
\begin{tabular}{ll}
$+\infty$                                    & $r_{ij} \leq d_{ij}$ \\
                                             &                     \\
$\sum_{i< j}^N \frac{q_iq_j}{\epsilon_0 r_{ij}}$ & $r_{ij} > d_{ij}$ \\
\end{tabular}
\right.
\label{hamilt1}
\end{equation}
here  $r_{ij} = \mid {\bf r_j-r_i}\mid$ is the distance of ions $i$ and $j$. In what follows, charges are expressed in atomic 
units, i.e., in units of the (positive) electron charge, while distances will be expressed in units of the cation diameter 
$d_{++}=1$. For any extended system, represented by periodically repeated finite ($N$ ions) samples, overall charge neutrality 
($\sum_{i=1}^N q_i=0$) is strictly required to have a finite electrostatic energy density. For the finite spherical systems that
are investigated below, samples may have a non-vanishing net charge $Q=\sum_{i=1}^N q_i$. Again for the sake of simplicity, the 
simulation procedure is described as it applies to a mono-valent MX salt consisting of cations and anions of the same charge 
$\mid q_+\mid=\mid q_-\mid= 1\ e$. This restriction is not essential and can be easily removed.

The Hamiltonian  in Eqs.~\ref{hamilt0},-\ref{hamilt1} is a qualitative but nevertheless relevant model for water solutions of 
alkali halide salts, and sometimes it has also been used for slightly more complex systems, such as alkaline-earth halide 
(MX$_2$) salts\cite{mgcl2} in water. To lend some real-life connection to the model, its scaled variables can be transformed to 
real ones using the following arguments. Based on the evidence from electrochemistry\cite{bockris} and also from recent 
microscopic measurements,\cite{rigo} cations consists of the bare metal ion accompanied by its hydration shell. The anion, whose
hydration shell is far less rigidly bound, consists primarily of the bare halide ion. Because of this assumption, the radius of 
cations and anions in somewhat asymmetric NaCl or KCl / water solutions is nearly the same, and, following 
Ref.~\onlinecite{valleau}, the common radius is set to $d=4.25$ \AA\ , which applies to cation-cation, cation-anion and 
anion-anion pairs. Strictly speaking, this cation-anion symmetry defines the so called {\it restricted} primitive model, but in 
the present case, the equality or additivity of the distances of closest approach are only accidental, and no essential change 
would be required if cations and anions had different diameter, valence (i.e., charge) and relative density. In what follows, 
all simulations are carried out at $T=298$ K, temperature at which the dimensionless coupling parameter 
$\beta^{\ast}=e^2/(\epsilon_0 d_{++} k_B T)$ is equal to $1.609$, which corresponds to low Coulomb coupling. Moreover, the packing
fraction $\hat{e}=\frac{\pi \rho}{6} \left[d_{++}^3+d_{--}^3\right]$ of the simulated systems is also low ($5\ 10^{-4} \leq 
\hat{e} \leq 0.1$), corresponding to relatively dilute electrolyte solution of concentration from $0.01$M to $2$M. In the 
expression defining $\hat{e}$, $\rho$ is the common density of cations and anions. Both conditions of low Coulombic and packing 
couplings are needed for the application of grand canonical Monte Carlo in the present implementation of the method.

Since water is only implicitly present, the pressure of the particles represents the osmotic pressure only. This might make the
samples too compressible to be meaningfully simulated in the NPT ensemble, and the essential inhomogeneity and anisotropy of the
interface add to this difficulty. Hence the connection to a precise thermodynamic state is established by using the GCMC 
formalism, as already done nearly half a century ago for the same model in Ref.~\onlinecite{valleau}. In this scheme, the normal
Monte Carlo protocol to update the particles' positions and to sample the phase space of $N$ ions is supplemented by {\it moves}
that attempt to change the number of ions in the system. Normal and grand-canonical moves alternate each other at random, with 
only the ratio of their relative probability being specified. To keep constant the overall charge in the sample, cations and 
anions are added or removed in neutral pairs. Since at the simulation conditions the salt molecules are dissociated, the 
addition of cations and anions occur independently at random positions within the space available to the ions. Also, the cation 
and anion whose removal is attempted are selected at random among all the ions in the sample. The acceptance probability for 
addition and removal of ion pairs is easily derived from the analytical expression of the grand canonical partition function for
a system at temperature $T$ and chemical potential of the neutral MX molecules $\mu=\mu_++\mu_-$. Following again 
Ref.~\onlinecite{valleau}, the addition of a neutral MX salt molecule, increasing by one the number of cations ($N_+ 
\rightarrow N_++1$) and anions ($N_- \rightarrow N_-+1$), is accepted with probability: 
\begin{equation}
f_{i\rightarrow j}=\frac{1}{(N_++1)(N_-+1)} \exp{\left[ \beta \mu -\beta (U_{N+2}-U_N)\right]}
\end{equation}
In a related way, removing one MX salt molecule, thus decreasing by one the number of cations ($N_+ \rightarrow N_+-1$) and 
anions ($N_- \rightarrow N_--1$), is accepted with probability:
\begin{equation}
f_{i\rightarrow j}=N_+ N_- \exp{\left[-\beta \mu +\beta (U_{N}-U_{N-2})\right]}
\end{equation}
As already stated, this approach works satisfactorily only for dilute systems, such as the implicit solvent electrolytes 
investigated here. This restriction is difficult to remove, but a promising approach to treat concentrated electrolytes such as 
ionic liquids will be briefly discussed in the last section of this paper.

In the series of simulations described below, the independent thermodynamic variable that uniquely defines the state of the
electrolyte solution is the excess chemical potentials in $k_BT$ units $\beta (\mu-\mu_{ideal})$ (here $\beta=1/k_B T$), which, 
for a monovalent MX salt, equals $2\langle \ln{\gamma_{\pm}}\rangle$, where $\gamma_{\pm}$ is the mean activity 
coefficient.\cite{valleau} In a
preliminary stage of the computation, this quantity has been determined by GCMC as a function of the molarity M of the 
electrolyte solution, then used in the following stages to fix the thermodynamic state of the bulk system in equilibrium with 
the inhomogeneous electrode / electrolyte sample.

The second crucial element in the model electrochemical cell is the metal electrode, able to transfer electrons between the two 
sides of the interface. The simplest extension of the primitive electrolyte model to investigate the electrified interface 
consists of a system in which the charged hard sphere ions are confined between two structureless planar parallel surfaces 
impenetrable by the ions, and carrying a surface charge density $\sigma$ in $[e/d_{++}^2]$ units. This paradigmatic model has 
been simulated several times in the past (see, for instance, Ref.~\onlinecite{torrie, tosi, squires}), providing a wealth of 
information on simple electrified interfaces as a function of electrolyte concentration and surface charge density, including 
the electrostatic potential drop across the interface, the differential capacitance, the density profile of the electrolyte ions
along the direction perpendicular to the planar surface, the specific adsorption of ions on the metal electrodes. The scheme, 
however, has one drawback, since necessarily it introduces two electrodes and two interfaces, which, away from the PZC, may 
become inequivalent, unless the electrodes and ions are both symmetric. Given the (still) microscopic sizes of the simulated
samples, disentangling the properties of the two electrified interfaces might become uncertain. In real-life, macroscopic 
systems, both interfaces are in equilibrium with a common bulk, and, although related, do not interact directly with each other.
Ideally, it would be advantageous to deal with half of the electrochemical cell at a time, which is in fact a conceptual 
idealisation widely used in electrochemistry. For this reason, we resort to a variant of the basic model in which the 
electrolyte is enclosed into a spherical surface of radius $R_e$ which plays the role of the unique electrode in the half cell. 
Then, the state of the electrolyte is identified and kept stationary in time by the GCMC approach, offsetting the effect of the 
finite sample size, and also of the electrostatic bias and electric current flowing through the electrode. Needless to say, the 
disadvantage is that the single-electrode model introduces a curvature dependence of the results, which however, can be 
estimated and compensated. This aspect will be discussed in the following sections and in the Supplementary Information (SI) 
document.

The finite and inhomogeneous samples under investigation will be simulated under charge control, as opposed to other methods 
that describe systems under electrostatic potential control,\cite{linnea, constphi} whose comparison with experiments might be 
more direct. However, the electrostatic properties of the curved interfaces simulated in the present study, such as the surface 
charge, the electrostatic potential drop across the interface, the screening length and the interfacial capacitance, have been 
quantitatively determined, in such a way that the results obtained under charge-control conditions can be translated into the 
potential-control picture.

To mimic real systems, the surface charge density on the spherical electrode would exactly balance the overall charge of the 
electrolyte inside the spherical cavity:
\begin{equation}
\sigma=\frac{-\left(\sum_{i=1}^N q_i\right)}{4 \pi R_e^2}
\end{equation}
In the single electrode model, the non-vanishing $\sigma$ does not play an explicit role, but it is implicitly accounted for by 
introducing a charge imbalance on the electrolyte side of the interface. This initial charge is left unchanged by the GCMC that 
adds or removed neutral pairs of ions. Because of Gauss's theorem, the correct electric field condition $E_r=4 \pi \sigma$ at 
the interface is automatically satisfied, provided the charge density distribution is, on average, spherically symmetric both on
the metal and the electrolyte sides of the interface. This observation reflects the fact that the electric field due to a 
homogeneously charged spherical surface identically vanishes on its inside.

It is important to remark that the electrostatic field conditions at the interface are not rigorously treated, even at the
simplest (perfect conductor) level of image charge. However, image charges can easily be computed for spherical conducting
interfaces\cite{feynman} and their effect can be accounted for by a slight extension of the model. 

The novel aspect of the present model consists of a new type of move that introduces the electron transfer across the metal /
electrolyte interface, as well as a protocol intended to match this process with the underlying diffusion of the ions that keeps
the system in a stationary state, despite the change of ions' concentration in the immediate vicinity of the electrode driven by
the electron transfer itself. 

As stated in the introduction, we will be looking at cations which may closely approach the cathode, picking up an electron thus
reducing their positive charge by one unit. The simplest model for the electronic part of the process, in which electrons move 
from the metal valence band to ionic/atomic states, corresponds to a stochastic surface hopping picture.\cite{tul, hop} This 
assumes that the quantum mechanical electron jump takes place instantaneously, after a latency representing the time needed to 
overcome the free energy barrier between the initial and final state (see below), and accounting also for the probabilistic 
nature of quantum transitions. The kinetics of the process corresponds to that of first order chemical reaction,\cite{atk} whose
rate is proportional to the abundance of ions of the reactive species (cations, in this case) located within a distance 
$\delta R$ from the metal surface such that the transition probability is not vanishingly small.

The present model characterises the inherent kinetics of the electron transport through the interface by the parameter $k_e$, 
expressing the rate of the electron transfer from the cathode to a single active cation. The aim of the model is to determine 
how this basic rate combines with all the other many-ion effects to produce the flow of charge through the system. This output
quantity, which, in the following, will also be referred to as the {\it overall electron transfer rate}, has the dimension of a 
current density, being expressed as charge transferred per unit area and unit time. 

The $k_e$ parameter can be computed by quantum perturbation theory,\cite{schmic} provided the coupling of the two sides is weak,
or by many-body methods based, for instance, on the Anderson-Newns Hamiltonian.\cite{an1, an2} For the sake of simplicity, in 
what follows, it is assumed that the top of the electrode valence band has {\it sp} character, hence the corresponding valence 
density of states is broad ({\it wide band assumption}) and below the Fermi level it changes slowly with energy. Then, despite 
the fact the cation has an energy spectrum of sharp levels, the rate will depend weakly on the bias, and will not display the 
resonance spikes expected for electrodes made of transition metals. Instead, the dependence of the transfer rate from the 
metal-ion separation, which could also be computed by the same electronic structure methods, is strong, as suggested from the 
fact that, according to perturbation theory, $k_e$ depends on the superposition of the electronic wave functions centred on the 
metal and on the ions, which decays exponentially with increasing their mutual distance. Hence, the model will only include the 
charge transfer for the ions whose radial distance $\delta R$ from the electrode is of the order of the \AA\ . In principle,
$k_e$ accounts for electronic effects only, resulting from the activation (free) energy $\Delta G$ of electronic origin, that
characterises the kinetics of the reaction between the metal and the bare cation, which might be seen as a tunneling process. 
However, in schematic models like the present one, other effects  might be folded into the definition of $k_e$. For instance, 
the electron transfer often involves a positive {\it reorganisation energy} $\lambda$, which is the free energy barrier 
associated with the loss (or, at least, drastic reorganisation) of the tightly bound hydration shell of the cation being 
neutralised. Hence, the total activation energy $E_{act}$ that is responsible for the $k_e$ determination 
is:\cite{marcus, marcusb, hush}
\begin{equation}
E_{act}=\frac{(\lambda+\Delta G)^2}{\lambda}
\end{equation}
which may easily reach and surpass the eV energy scale. Hence, in most cases the electron transfer is a relatively slow process,
whose rate per ion typically will be below the inverse nanosecond. The time step of atomistic molecular dynamics simulations is 
of the order of the femtosecond, implying that more than $10^6$ MD updates of forces, velocities and positions are needed per 
every event (on average) on each particle within the jump distance. Of course, a large interface will involve many active 
particles, whose number grows like the interfacial area, increasing the rate of statistics accumulation. Since, however, the 
computational cost per step grows approximatively like the square of the  number of particle, increasing the sample size does 
not provide a scaling advantage.

To account for all these aspects in the simplest possible way, the $k_e$ in the present study is a free parameter to be varied
in order to investigate the model properties. Moreover, in the present model, the same rate constant $k_e$ is attributed to all
cations located within a distance $\delta R=d_{++}/2$ from the metal surface. Each cation satisfying this distance condition has
the following probability to be still unreacted at time t:
\begin{equation}
p_{survival}(t)=\exp{ \left[ -k_e t\right]}
\end{equation}
For any given cation residing in the active region, the time to (first) neutralisation is a random variable $\Delta t$ whose 
distribution $p(\Delta t)$ is:
\begin{equation}
p(\Delta t)=k_e \exp{ \left[ -k_e t\right]}
\label{decay}
\end{equation}
In this way, the lifetime of a cation adsorbed on the electrode within the distance $\delta R$ turns out to be $\tau=k_e^{-1}$.
Simulating this process is the same as simulating the radioactive decay of an assembly of metastable nuclei, or the 
de-excitation of metastable electronic excited states. In these analogous cases, the probability distribution in Eq.~\ref{decay}
is interpreted as describing the first escape time out of a metastable minimum. This same decay/relaxation picture is 
reminiscent of the way kinetic Monte Carlo simulates the system escape from any given free energy basin of the system under 
consideration.\cite{KMCb} In a broader context, the kinetic approach to describe all these processes are based on the theory
of the master equation.\cite{master} A variety of algorithms\cite{gilles} and also computer codes are available for this type of
simulations.

The crucial aspect of the method that is proposed below is the matching of the stochastic charge transfer process with the 
second dynamical process, i.e., the self-diffusion of the electrolyte ions, that is needed to keep the system in a stationary 
state. The problem addressed by the discussion that follows is how to join these two components of the dynamics, mixing them in 
the right proportion that reproduces the mutual effects they have on each other. The electron transfer itself already has its 
own real time clock, whose beat is given by $k_e^{-1}$. Then, the idea is to use the ions' diffusion coefficient computed or 
measured in real time to set up a time scale for the response of the electrolyte density distribution to the disturbance due to 
the flow of charge through the system.

Then, the synchronisation of the stochastic transfer of electrons at the interface with the Brownian dynamics of ions in the 
electrolyte solution is achieved through the following steps. First, the mean square displacement per ion $\langle \mid 
{\bf r_i}(\tau+\tau_0)-{\bf r_i}(\tau_0)\mid^2\rangle_{MC}$ as a function of the time interval $\tau$ is evaluated for all the 
samples of interest. Here $\tau$ is the MC time, measured in MC steps, and {\it step} consists of a complete sweep of attempted 
MC moves over all ions. To be precise, following standard practice and for the reasons discussed below, the ions whose 
displacement is attempted are selected at random. Therefore, the {\it sweep} term will only be used as a concise way to denote
a number of attempted moves equal to the number of particles. It will not imply, in particular, that during a sweep each ions
will be tested for displacement, or, much less, that each atom is displaced only once or event at all during a sweep.
In the definition of the average displacement, the $\langle ... \rangle_{MC}$ notation indicates averaging over the initial time
$\tau_0$ and also over all ions of the sample. 

In a fluid system, the Metropolis Monte Carlo dynamics results in a diffusive regime in which the mean square 
displacement of particles asymptotically grows linearly with MC time $\tau$. Then, the self-diffusion coefficient $D_{MC}$ of 
ions in MC time is defined through the same Einstein's relation that defines the real time diffusion $D$:
\begin{equation}
D_{MC}=\frac{1}{6} \lim_{\tau \rightarrow \infty}
\frac{\langle \mid {\bf r_i}(\tau+\tau_0)-{\bf r_i}(\tau_0)\mid^2\rangle_{MC}}{\tau}
\end{equation}
Needless to say, $D_{MC}$ depends on the electrolyte concentration in solution, and on the step $\Lambda$ of the attempted MC 
single-ion moves along each Cartesian direction, in analogy with the real-time diffusion coefficient $D$ which depends on the 
electrolyte concentration and on the mass of the diffusing particles. Notice that for the symmetric electrolyte model, only one 
diffusion coefficient is needed. The discussion and ensuing simulation protocol could easily be extended to cover the more 
realistic case of an asymmetric electrolyte, which will require two different steps for cations and anions to reproduce their 
two distinct diffusion constants.

The correspondence between real ($\delta t$) and MC ($\delta \tau$) time is established by stating that the $\delta t$ and 
$\delta \tau$ intervals are equivalent if the increase of the mean square displacement per ion is the same in the two cases. 
Starting from the Einstein's relation, it is easy to verify that this condition is equivalent to impose that:
\begin{equation}
\delta t=\delta \tau \frac{D_{MC}}{D}
\end{equation}
where $\delta t$, $\delta \tau$, $D$, and $D_{MC}$ are expressed in their own units, for instance, nanoseconds, MC steps,
\AA$^2$/ns and \AA$^2$/MC step, respectively.

Then, the knowledge of the real time diffusion coefficient for the same system that is being investigated by MC is required to
define the correspondence between the MC time and real time. Ideally, one would use the real-time diffusion coefficient computed
by molecular dynamics for the same model investigated by MC. Since, in the present case, the implicit-solvent, hard sphere model
is not likely to have a real-time diffusion constant comparable to those of alkali halide solutions, we use, or, better, 
approach the experimental self-diffusion constants of Na$^+$ and Cl$^-$ reported in the literature as a function of temperature 
and electrolyte concentration. The diffusion properties of Na$^+$ and Cl$^-$ will be symmetrised by introducing the average 
diffusion coefficient  $D=\sqrt{D_+ D_-}$. Starting from these considerations, the protocol used to relate the real and MC time 
is fully defined in the Results sections, where diffusion coefficients and computational parameters are listed and discussed. 

In what follows, $\delta \tau$ will represent the unit MC time (i.e., one MC step), while $\delta t$ will represent the
corresponding real time interval, expressed, in the present case, in ns. In other terms, $\delta t$ is the real time covered by
a set of $N$ attempted MC moves, where, as before, $N$ is the number of ions. As already stated, for any given electrolyte 
concentration, the diffusion 
coefficient in MC time depends on the amplitude $\Lambda$ used, along each Cartesian coordinate, for the MC attempted single ion
displacement. In dilute systems, whose dynamics is virtually Brownian, $D_{MC}$ will scale nearly quadratically with this step 
amplitude. Slight deviations from this simple law are due to the dependence of the MC acceptance / rejection probability on 
$\Lambda$ (see Sec.~S1 in SI). The overall near quadratic dependence gives a suitable handle to tune $\delta t$: increasing
$\Lambda$ increases $D_{MC}$, and thus increases $\delta t$ by the same nearly quadratic ratio. If $\delta t$ is too short, the 
advantage of the MC dynamics with respect to molecular dynamics is reduced. If $\delta t$ is too long, the time resolution of
the method is low, and might introduce artifacts in the combined electron transfer versus ionic diffusion dynamics. 

It might be worth emphasizing that the full sweep of $N$ attempted single-particle MC moves has been introduced as a unique and 
indivisible unit of time. The reason of this choice is that in this way $\delta \tau$ is certainly long enough to allow the 
system to lose memory of its previous state at each step, fulfilling all requests of Markovian evolution and purely diffusive 
behaviour. However, as already said, particles are not moved in a regular order, but each attempted move is applied to an ion 
chosen at random. Hence, during one $\delta \tau$ (i.e., what we called a {\it sweep}), it might happen that a given ion is 
moved more than once, and certainly many ions do not move even once. The random choice of the ion makes the flow of time more 
uniform, and allows to split the unit of time $\delta \tau$ in sub-units, corresponding to a fraction $n$ of the $N$ attempted 
moves. This might be needed to increase the time resolution at low ion density, when diffusion is so fast to correspond to long 
$\delta t$ values. This choice, which is less rigorous than the previous one since it does not fully guarantee the loss of 
memory from one step to the next, is nevertheless exploited in the computations presented in Sec.~\ref{results}.

Taking into account all these aspects, the MC simulation will proceed according to the following block diagram:

\begin{enumerate}
\item  Electron transfer simulations are started from one configuration selected at random from the GCMC simulation of the 
 sample, and the system time $t_s$ is initialised to zero. All cations in contact with the electrode (say, within a skin 
 distance $\delta R$) are initialised with a label (i.e., a binary state variable) which says that they are {\it active}, i.e.,  in the process of negotiating the charge transfer, whose rate, per  cation, is $k_e$. Also, a random variable $t_e$ is 
 associated to each of these active ions, drawn from the exponential distribution $p(t)=k_e \exp{[-k_e t]}$. The $t_e$ 
 variable, different for each cation, gives the first passage time for their change of oxidation state, which, in what follows, 
 will also be indicated as the ion {\it expiration} time. In other terms, this random variable tells when  each ion will change 
 state, provided it remains active in the meantime. This expiration time is not initialised for inactive particles.
\item The simulation is carried out by performing a MC step consisting of $N$ attempted displacements (or over a subset of $n$ 
attempted displacements) of ions chosen at random.  At the end, the system time is advanced by the corresponding real time 
$\delta t$ (or $n \delta t/N$). 
\item The expiration time of each {\it active} ion is compared with the new time. If the accumulated system time is longer than
 the expiration time, the ion undergoes the charge transfer, carried out as detailed below. If the accumulated time is shorter
 that the expiration time of the given ion, this one will continue accumulating time as an active ion.
\item New ions might have entered the active range during the diffusion stage. Their binary state variable is turned to 
 {\it active} For each of them, a random variable $t_r$ is drawn from the exponential probability distribution, and their 
 expiration time is set to $t_s+t_r$.
\item Steps 2, 3 and 4 are repeated until the accumulated system time exceeds the preset value.
\end{enumerate}

In applying this algorithm, the following simplifying assumptions are made about the system and the electron transfer process. 
Since the electrode/electrolyte model is very idealised, and does not target any specific system, both the electron transfer 
rate and its distance dependence will be decided a priori, without resorting to an electronic structure computation. In
particular, the rate will be the same for all ions of the reactive type (cations, in the present case) located within one ionic 
radius  ($\delta R=d_{++}/2)$ 
from the metal electrode surface. The value of the rate per active ion will be used as a parameter to simulate low-flow (0.01 
ns$^{-1}$) / high-flow (0.1 ns$^{-1}$) conditions. The attempted electron transfer is accepted a priori, without any 
acceptance/rejection stage, based, for instance, on the energy change. This is because the scale of the electron energy change 
is supposed to be much higher than thermal energy, hence the process is driven by the electro-motive force exerted on the half 
cell by the external circuit.

\begin{figure}[!htb]
\begin{minipage}[c]{\textwidth}
\vskip 0.7truecm
\begin{center}
\includegraphics[scale=0.50,angle=-0]{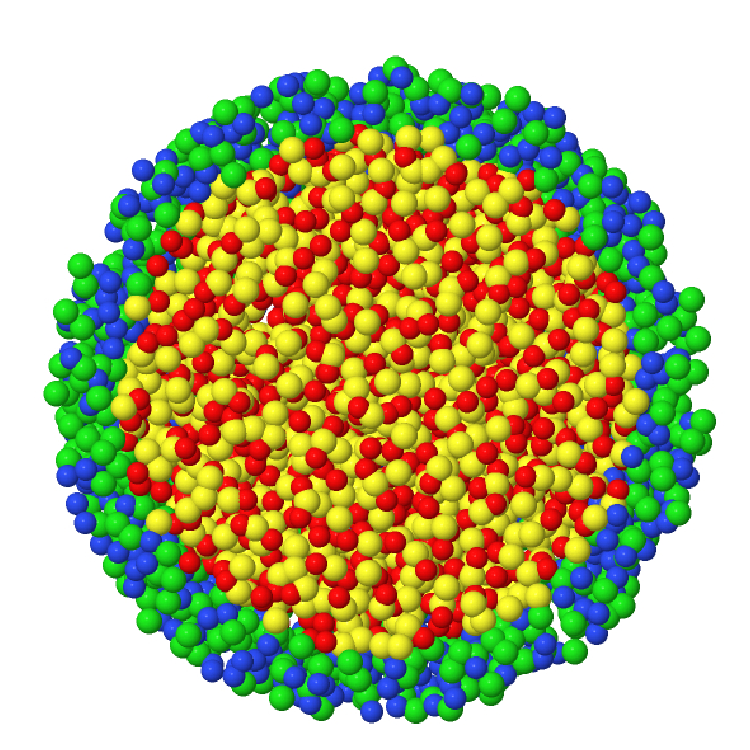}
\vskip 2.0truecm
\caption{Snapshot of the smallest ($N=4000$) and most concentrated ($2M$) sample.
Red and yellow dots represent cations and anions, respectively, in the inner sphere, accounting for $51.2$ \% of the volume; 
blue and green dots represent cations and anions, respectively, in the outer sphere, accounting for the remaining $48.8$ \% of 
the volume. The radii of the ions are not on scale with the radius $R_e$ of the spherical electrode. The target molarity is 
tuned on the inner sphere only, but the GCMC addition and removal of ion pairs is carried out over the entire volume (see 
Sec.~\ref{method}.
}
\label{spheres}
\end{center}
\end{minipage}
\end{figure}

In the present model, no detailed chemical reaction is specified for all the ions during the flow of current through the half 
cell. This aspect might be added at a later time to extend the scope of the model and to apply it to a specific problem. Right
now, the change that the active ions undergo upon changing their oxidation state is used to model some additional condition on 
the system evolution. In the present study, the electron transfer mechanism has been tuned in order to drive a current density
${\bf j}({\bf r})$ through the systems whose radial component $j_r ({\bf r})$ is, on average, uniform over the electrode cavity.
Of course, {\it radial} refers to the spherical polar coordinate system concentric with the spherical electrode. This request 
has been made to mimic the homogeneous distribution of current density at planar interfaces, and has been fulfilled in the 
following way. First, the cation which is undergoing neutralisation is removed, i.e., turned into a neutral particle which is 
only implicitly represented by the model (like the solvent molecules). Instead, a cation is randomly inserted at a new random 
position ${\bf r}_n$ within the volume enclosed by the spherical electrode. The new position ${\bf r}_n$ is selected with 
probability $p_r(r_n)$ which is not uniform, but, ideally, depends on $r_n=|{\bf r}_n|$ as $1/r_n$. In this way the radial 
current density is predicted to be the same through all spherical surfaces concentric with the electrode. However, the 
probability $p_r(r_n)$ is singular and non-integrable at the origin. To avoid this problem, the (unnormalised) probability 
distribution $p_r$ is slightly modified into:
\begin{equation}
p_r(r)=\frac{1}{1+r}
\label{constj}
\end{equation}
and it has been verified that in this way the radial component of ${\bf j} ({\bf r})$ is virtually uniform over the whole 
sample (See the Results section). If, in the new location, the inserted cation is superimposed to any other ion, a new random 
position is selected, until the move is successful. In this sense, as we already wrote, the move occurs a priori.

In a different scenario, a fraction $0 \leq \nu \leq 1$ of the neutralising ions is exchanged with an anion chosen at random 
from the entire anion population. The remaining $(1-\nu)$ fraction of neutralising ions will undergo the same relocation 
process describe before. Both processes are weighted with the probability $p_r$ defined in the previous sentence, applied
to the new position of the interchanged/relocated cation, or to the old position of the interchanged anion. This combination of 
simple relocation and cation/anion exchange will result in the simultaneous presence of a cathodic and anodic current through 
the system, whose relative strength depends on $\nu$. Both currents turn out to be homogeneous across all spherical surfaces 
concentric with the electrode, as homogeneous is the net current in systems delimited by planar interfaces.

Two remarks are in order. First, the condition of (nearly) uniform radial current density $j({\bf r})$ over the whole sample 
implies a violation of the continuity equation, whose origin and impact are discussed in the Results section. Second, the 
transformation of the neutralising cation into a neutral particle only implicitly represented in the model is somewhat 
artificial. This choice has been made in the present study for the sake of simplicity only. In further developments of the 
method, the transforming ions could contribute to the grow of the electrode through an electro-deposition process, or could take
part in some more complex electrochemical reaction, opening the way to simulating Galvanic or fuel cells. The further redox 
reactions required to model these systems and phenomena could be treated in analogy with the electron transfer at the 
electrode/electrolyte interface.

\section{Results}
\label{results}

\begin{table}[ht]
\caption{Geometric, thermodynamic, structural and electrostatic properties of the neutral simulated systems. M: molarity
concentration of the electrolyte in the water solution in equilibrium with the spherical interface; N: {\it nominal} number of 
ions, see Sect.~\ref{prepare}; 
$R_e$: radius of the spherical interface;  $2 \langle \ln{\gamma}_{\pm} \rangle=\beta (\mu-\mu_{ideal})$: excess chemical 
potential of an ion pair. $\gamma_{\pm}$ is the mean ionic activity; $\Gamma_+=\Gamma_-$: adsorption coefficient of cation and 
anions expressed in number of ions per unit area ($d_{++}^2$), which are equal in the case of neutral electrode ($\sigma=0$); 
$\rho_+(R_e)=\rho_-(R_e)$: ionic number density at contact with the electrode, expressed in number of ions per unit volume 
($d_{++}^3$). The determination of these quantities from the raw simulation data is detailed in Sec.~S2 of SI.
}
\begin{center}
\begin{tabular}{|c|c|c|c|c|c|c|}
    \hline
M&$\lambda_{DH}/d_{++}$&N&$R_e/d_{++}$& $2 \langle \ln{\gamma}_{\pm} \rangle$&$\Gamma_+=\Gamma_-$ & $\rho_+(R_e)=\rho_-(R_e)$ \\
\hline
    \multirow{3}{*}{ 2 }& \multirow{3}{*}{1.043}& 4000 &17.28& 0.522 $\pm 0.003$&$(5.8\pm 0.4)\ 10^{-3}$&0.1247 $\pm 0.0004$  \\
                        &  & 8000 &21.78 & 0.524 $\pm 0.003$  & $(5.8\pm 0.4)\ 10^{-3}$ & 0.1262 $\pm 0.0004$   \\
                        &  &16000 &27.44 & 0.524 $\pm 0.003$  & $(6.3\pm 0.4)\ 10^{-3}$ & 0.1249 $\pm 0.0007$   \\
    \hline
    \multirow{3}{*}{ 1 }&\multirow{3}{*}{1.475} & 4000  &21.78& -0.252 $\pm 0.003$   & $(1.1\pm 0.2)\ 10^{-3}$ & 0.0509 $\pm 0.0002$\\
                        &  & 8000  & 27.44& -0.258 $\pm 0.003$   & $(0.88\pm 0.2)\ 10^{-3}$ & 0.0507 $\pm 0.0003$    \\
                        &  &16000  & 34.57& -0.258 $\pm 0.004$   & $(0.78\pm 0.2)\ 10^{-3}$ & 0.0510 $\pm 0.0003$    \\
    \hline
\multirow{3}{*}{0.5}&\multirow{3}{*}{2.086}  & 4000  &27.44& -0.492 $\pm 0.004$ &$-(3.5\pm 1)\ 10^{-4}$ & $0.02298 \pm 5\ 10^{-5}$\\
                        &  & 8000  &34.57& -0.496 $\pm 0.004$   & $-(4.2\pm 1)\ 10^{-4}$ & 0.02288 $\pm 5\ 10^{-5}$    \\
                        &  &16000  &43.55& -0.490 $\pm 0.004$   & $-(4.4\pm 1)\ 10^{-4}$ & 0.02293 $\pm 8\ 10^{-5}$    \\
    \hline
\multirow{3}{*}{0.1}&\multirow{3}{*}{4.665}&4000&46.92&-0.440 $\pm 0.004$&$-(2.6\pm 0.3)\ 10^{-4}$&$4.332\ 10^{-3}\pm 2\ 10^{-5}$ \\
                        &  & 8000  &59.11& -0.446 $\pm 0.004$   & $-(2.5\pm 0.2)\ 10^{-4}$ & $4.312\ 10^{-3} \pm 5\ 10^{-5}$    \\
                        &  &16000  &74.47& -0.447 $\pm 0.004$   & $-(2.8\pm 0.3)\ 10^{-4}$ & $4.376\ 10^{-3} \pm 2\ 10^{-5}$    \\
    \hline
\multirow{3}{*}{0.01}&\multirow{3}{*}{14.75}&4000&101.08&-0.1958 $\pm 0.001$&$-(2.8\pm 0.2)\ 10^{-5}$  & $4.451\ 10^{-4} \pm 2\ 10^{-6}$\\
                        &  & 8000  &127.35& -0.1970 $\pm 0.001$ &$-(3.1\pm 0.2)\ 10^{-5}$ & $4.486\ 10^{-4} \pm 3\ 10^{-6}$ \\
                        &  &16000  &160.45& -0.1945 $\pm 0.001$ &$-(3.0\pm 0.2)\ 10^{-5}$ & $4.501\ 10^{-4} \pm 1\ 10^{-6}$ \\
    \hline
\end{tabular}
\end{center}
\label{activity}
\end{table}

\subsection{Sample preparation and characterisation of the electrode/electrolyte interface}
\label{prepare}

At first, neutral samples have been prepared to represent spherical electrode / electrolyte solution interfaces whose fluid 
side is in equilibrium with electrolyte solutions of concentration (molarity) M=0.01; 0.1; 0.5; 1; and 2. To assess size (and 
therefore curvature) effects, each concentration has been simulated at three sizes, corresponding to $N/2= 2000$, 4000 and 8000 
neutral ion pairs, with $N$ being the number of ions in the sample. To be precise, these round $N$ values are the {\it nominal} 
numbers of ions in the sample, meaning that the sample volume contains precisely those numbers of ions at the target molar
concentrations. However, the actual average number might differ slightly because of fluctuations due to the grand-canonical MC 
method and because of the perturbation represented by the fluid/solid interface, that is measured by the adsorption coefficients
$\Gamma_+$, $\Gamma_-$ for cations and anions, respectively (see Tab.~\ref{activity}). All simulations have been carried out at 
$T=298$ K, with water being represented by a (relative) dielectric constant $\epsilon_0=78.5$.

The sample preparation consisted of the following steps. Since the molarity (M) times the Avogadro number ${\cal{N}}_A$ gives
directly the number density of molecules per liter of solution, the volume $V$ of each spherical sample is set to:
\begin{equation}
V=\left( \frac{N}{2}\right)\frac{1}{M {\cal{N}}_A}
\end{equation}
with the result being expressed in litres.

Each sample is progressively filled by grand canonical MC (see Sec.~\ref{method}), using the excess chemical potential as the 
independent variable that controls the average density, and thus the thermodynamic state of the sample. To be precise, the 
target density is imposed not on the total volume, but only on the central portion corresponding to $51.2$ \% of the volume (up
to 80\% of the radius), since the density of the external portion, accounting for the remaining 48.8 \% of the volume, might be 
slightly affected by the specific adsorption of ions at the electrode surface. This subdivision of the volume and of the 
particles it contains is illustrated in Fig.~\ref{spheres}. 

The primary thermodynamic quantity determined in these computations has been the mean ionic activity coefficient $\gamma_{\pm}$,
whose logarithm is related to the excess chemical potential of the two ion species according to the equation:
\begin{equation}
2 \langle \ln{\gamma_{\pm}}\rangle=\beta (\mu-\mu_{ideal})
\end{equation}
where $\mu\equiv \nu_+ \mu_+ + \nu_- \mu_-$, with $\nu_+$, $\nu_-$ giving the stoichiometry of the M$_{\nu_+}$X$_{\nu_-}$ 
electrolyte compound. As already stated, in the present case, $\nu_+=\nu_-=1$. The results of this sample preparation stage are 
summarised in Tab.~\ref{activity}.

\begin{table}[ht]
\caption{Properties of the electrostatic double layer as a function of electrolyte concentration, surface charge and size of the
sample. Slight variations in the surface charge for sample of different size are due to the fact that the charge is quantised 
in units of $e$. 
}
\begin{center}
\begin{tabular}{|c|c|c|c|c|}
    \hline
M & $R_e/d_{++}$ & $R_e$ [\AA\ ] & $\sigma d_{++}^2$ & $\Delta \varphi=\beta e \langle \varphi(R_e)-\varphi(0) \rangle$  \\
\hline
 \multirow{3}{*}{ 2 } &17.28&  73.44 &-0.05328 & -0.2905  $\pm 0.01$   \\
                      &21.78&  92.55 &-0.05336 & -0.3005  $\pm 0.002$  \\
                      &27.44& 116.61 &-0.05328 & -0.2636  $\pm 0.005$  \\
\hline
 \multirow{3}{*}{ 2 } &17.28&  73.44 &-0.10655 & -0.6516  $\pm 0.003$  \\
                      &21.78&  92.55 &-0.10673 & -0.5999  $\pm 0.002$  \\
                      &27.44& 116.61 &-0.10656 & -0.5767  $\pm 0.006$  \\
 \hline
 \multirow{3}{*}{0.1} &46.92&  199.39&-0.07205 & -0.3527  $\pm 0.004$  \\
                      &59.11&  252.22&-0.07243 & -0.3582  $\pm 0.001$  \\
                      &74.47&  316.52&-0.07231 & -0.3446  $\pm 0.006$  \\
 \hline
 \multirow{3}{*}{0.1} &46.92&  199.39&-0.01441 & -0.6870 $\pm 0.004$   \\
                      &59.11&  252.22&-0.01448 & -0.6885 $\pm 0.001$   \\
                      &74.47&  316.52&-0.01446 & -0.6865 $\pm 0.006$   \\
 \hline
 \multirow{3}{*}{0.01}&101.08&  429.58&-0.00312 & -0.2428 $\pm 0.001$   \\
                      &127.35&  541.24&-0.00312 & -0.2548 $\pm 0.001$   \\
                      &160.45&  681.91&-0.00312 & -0.2366 $\pm 0.005$   \\
 \hline
 \multirow{3}{*}{0.01}&101.08&  429.58&-0.00312 & -0.5210 $\pm 0.001$   \\
                      &127.35&  541.24&-0.00312 & -0.4994 $\pm 0.002$   \\
                      &160.45&  681.91&-0.00312 & -0.4951 $\pm 0.004$   \\
 \hline
\end{tabular}
\end{center}
\label{electros}
\end{table}

\begin{figure}[!htb]
\begin{minipage}[c]{\textwidth}
\vskip 0.7truecm
\begin{center}
\includegraphics[scale=0.85,angle=-0]{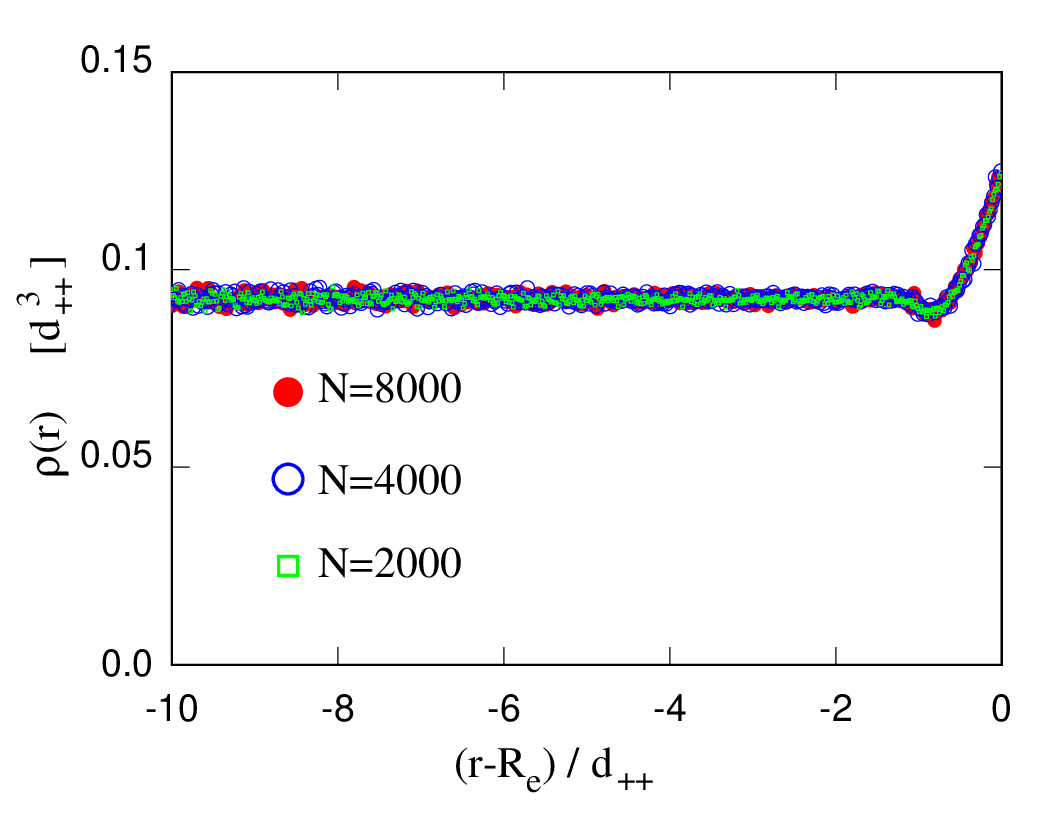}
\vskip 2.0truecm
\caption{Size dependence of the density profile for the neutral systems ($\sigma=0$) at 2M concentration.
}
\label{prozero}
\end{center}
\end{minipage}
\end{figure}

The main result coming from the data reported in Tab.~\ref{activity} is that $2 \langle \ln{\gamma_{\pm}}\rangle$ shows, as 
expected, a marked dependence on the electrolyte concentration, but a low dependence on sample size. This observation suggests 
that the protocol described above defines a unique bulk electrolyte solution in equilibrium with the interfaces of different 
curvature. Remarkably, the results agree to within the error bar with the results of Ref.~\onlinecite{valleau} despite the large
difference of sample size and length of the GCMC simulations, that reflects the vast improvement of computational equipment of 
the last 40 years. All these observations somewhat justify the usage of the spherical electrodes, and downplay the effect of 
curvature on the results.

\begin{table}[ht]
\caption{Inverse differential capacitance $C^{-1}_D$ (in $d_{++}$ units) as a function of system size and electrolyte 
concentration. $C^{-1}_D$ is computed by differentiating a linear interpolation of the  data for $\langle \varphi(R_e)-
\varphi(0) \rangle$ as a function of $\sigma$ using the data in Tab.~\ref{electros}.
}
\begin{center}
\begin{tabular}{|c|c|c|c|c|c|}
    \hline
  N     &   2M     &     1M      &   0.5M     &    0.1M   &    0.01M   \\
\hline
  4000  &  5.72    &    10.76    &  18.96     &   47.67   &   167.23   \\
  8000  &  5.62    &    11.84    &  18.68     &   47.54   &   160.02   \\
 16000  &  5.41    &    11.43    &  18.24     &   47.47   &   158.92   \\
\hline
\end{tabular}
\end{center}
\label{incapac}
\end{table}

\begin{figure}[!htb]
\begin{minipage}[c]{\textwidth}
\vskip 0.7truecm
\begin{center}
\includegraphics[scale=0.85,angle=-0]{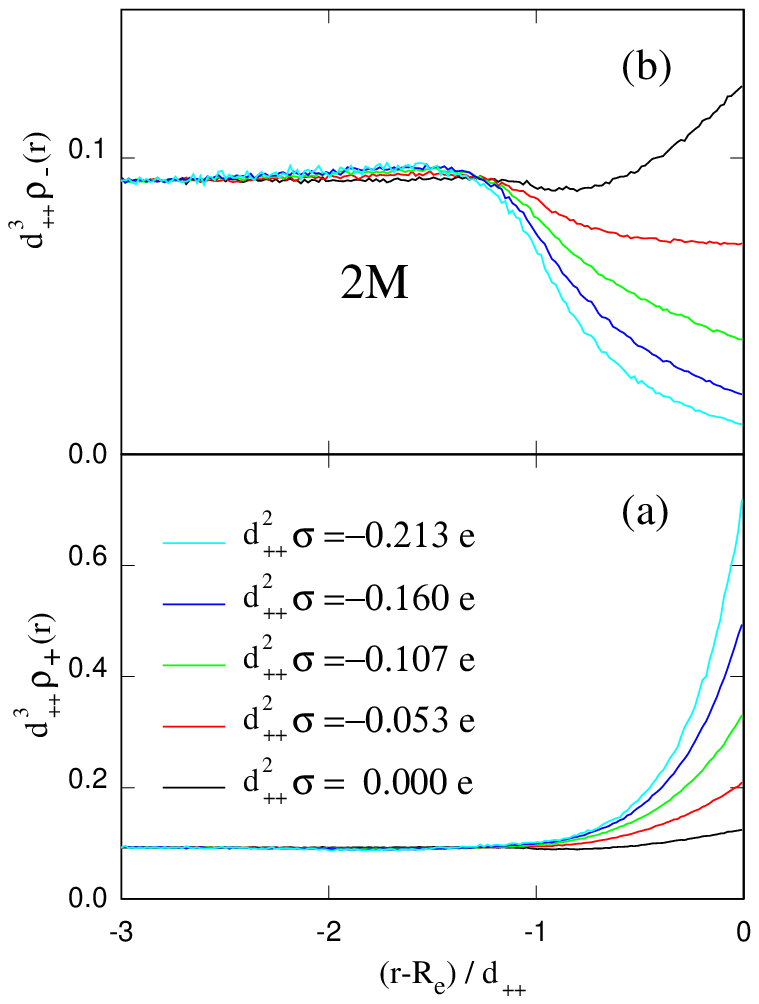}
\vskip 2.0truecm
\caption{Dependence of: (a) cation and (b) anion density profile as a function of the surface charge density $\sigma$.
Notice the change of scale between panel (a) and (b). Electrolyte concentration of the equilibrium bulk phase: 2M.
}
\label{rhoq}
\end{center}
\end{minipage}
\end{figure}

Then, simulations have been extended to charged samples, starting from the neutral ones, and changing a priori the $+/-$ charge 
of the ions until reaching the target charge imbalance, at the same time tuning the overall chemical potential to conserve the
$2 \langle \ln{\gamma_{\pm}}\rangle$ value of the sample. The slight tuning is needed because, at equal volume, neutral and 
charged samples contain somewhat different numbers of particles, hence the ideal contribution to the overall chemical potential 
(entering the GCMC acceptance probability) is slightly different. After this short calibration, the grand canonical part of the 
simulation proceeds as before with the attempted addition and removal of neutral ion pairs that maintain the total charge of the
sample. The results of these preliminary GCMC simulations for neutral and charged samples have been analysed to collect data on 
the structural and thermodynamic properties of the interface, and also on the dependence of these properties on the surface 
curvature and surface charge density $\sigma$. Here we only summarise the features that are relevant for the conductivity 
simulations. Further results, such as the electrostatic screening length and interfacial capacitance, are collected and briefly
discussed in Sec.~S3 of the SI document.

For what concerns the electron transfer process and the ensuing electrical conductivity through the system, the major structural
information from the preliminary simulation stage is given by the number and charge density profiles of the ions close to the
metal surface, since these determine the number of active ions and thus the observed rate of electron transfer. First, it has 
been verified that, over the range of sizes covered by the present study, the density profiles are nearly independent from the 
curvature of the metal electrode (See Fig.~\ref{prozero}).

Then, the dependence of the ionic density profiles on the surface charge density $\sigma$, has been analysed, focusing the
attention on their properties in the immediate vicinity of the electrode. As already repeatedly stated, the overall state of the
system is fixed by imposing the $2\langle \ln{\gamma_{\pm}}\rangle$ value of the whole spherical sample. Then, charged and 
neutral samples belong to the same thermodynamic state if they are in equilibrium with the same (neutral) bulk, identified by 
the value of $\beta (\mu-\mu_{ideal})=2\langle \ln{\gamma_{\pm}}\rangle$. It turns out that, if this condition is satisfied, the
molarity of the inner sphere (as before, the central 51.2\% of the spherical volume) is the same to within the error bar, 
confirming that the perturbation due to the surface charge is quickly screened by the redistribution of the ionic density 
profiles. Simulations have been limited to a rather narrow range of surface charge densities ($\mid \sigma \mid d_{++}^2\leq 
0.2$), which are the most relevant in electrochemical devices. The results are reported in Tab.~\ref{electros} and 
Fig.~\ref{rhoq}, and discussed in more detail again in Sec.~S3 of the SI document.

At a planar interface, the contact density depends on the surface charge according to the exact relation\cite{joel} expressing 
mechanical equilibrium normal to the interface:
\begin{equation}
k_B T \sum_{\alpha} \rho_{\alpha}(R_e)=P+\frac{2\pi \sigma^2 }{\epsilon_0}
\label{contc}
\end{equation}
where $P$ is the pressure of the bulk solution in equilibrium with the interface, $\alpha=+,-$ labels the two ion species, and 
$\sigma$ is the surface charge. The last term, in particular, can be identified as the pressure of the electric field at the
interface as given by the Poynting vector.\cite{jackson} Hence, for a neutral interface, the value of the electrolyte density 
(translated
in kinetic pressure units) at contact tends to the bulk osmotic pressure, apart, possibly, from a correction due to the surface 
curvature. The value of this contact density, always for neutral interfaces, is listed in Tab.~\ref{activity}. While we did not 
try to analytically derive the curvature dependence of the contact density, it has been verified that this dependence cannot be 
detected in the simulation results of Tab.~\ref{activity}, despite long runs and small statistical error bars. Moreover, the 
simulation results for neutral interfaces show that the contact between the electrode and the electrolyte turns from weakly 
non-wetting to wetting with increasing molarity M (See Sec.~4 of SI), reflecting the change of the excess pressure (i.e., the 
deviation from the ideal gas value $k_B T \rho$) from negative to positive. Needless to say, this model results might not hold 
for real systems, since it refers to the osmotic pressure only, and disregards the essential role of water in electrolyte 
solutions.

Of particular interest for the conductivity simulation is the dependence of the contact density on the surface charge. According
to Eq.~\ref{contc}, the sum of cation and anion densities at contact grows quadratically with the surface charge. The simulation
data for the charged interfaces satisfy this relation to within the error bar (See Fig.S6 in Sec.~4 of SI). The same data also
show that, as soon as $\mid \sigma\mid $ is not very low ($\mid \sigma \mid \geq 0.1$), only the majority species contributes 
significantly to the contact density. In the case of the simulated samples, the ions participating in the redox activity are the
cations, whose contact density as a function of $\sigma$ displays a marked asymmetry with respect to the $\sigma=0$ state, 
which, by definition corresponds to the potential of zero charge. The $\rho_+(R_e)$ dependence on $\sigma$, in particular, will 
be nearly quadratic when $\sigma << 0$, while it will decrease only slowly with $\sigma$ increasing above zero. This behaviour 
will be verified and further discussed in the following subsection.

A second quantity that might greatly affect the electron transfer process at the interface is the electrostatic potential drop 
across the interfacial double layer, which, assuming that the charge distribution is (on average) spherically symmetric, is 
given by:
\begin{equation}
\langle \varphi(R_e)-\varphi(0)\rangle=\frac{Q}{R_e}-4\pi\int_0^{R_e} r \rho_Q(r) dr
\label{integ}
\end{equation}
where $R_e$ is the radius of the spherical cavity enclosing the electrolyte solution. Collecting statistics for $\rho_Q(r)$ on
a grid gives a practical way to compute $\langle \varphi(R_e)-\varphi(0)\rangle$ by a simple 1D integration, which, however, is 
necessarily affected by the error of discretising $\rho_Q(r)$ on a grid. An approach which does not suffer from this limitation
is easily obtained. Since:
\begin{equation}
\rho_Q(r)=\langle \hat{\rho}_Q({\bf r} \rangle=\langle \sum_{i=1}^N q_i \delta(\mid {\bf r_i}\mid -r)\rangle
\end{equation}
one has:
\begin{equation}
\varphi(R_e)-\varphi(0)=\frac{Q}{R_e}-\langle  \sum_{i=1}^N \frac{q_i}{r_i}\rangle
\end{equation}
upon interchanging the integration and the averaging operations.

Therefore, the electrostatic potential drop can be computed by averaging on the fly $\sum_{i=1}^N \frac{q_i}{ r_i}$, where $r_i$
is the radial component of the position vector in a spherical coordinates systems whose origin is the centre of the spherical 
cavity. The $\sum_i q_i/r_i$ estimator has the obvious problem that it is bound neither from below nor from above, since ions 
can approach the origin without limitations. Hence, the variance might be very large, although the probability of very high or 
very low contributions to the estimator is vanishingly small, being weighted by the $r^2 dr$ volume element. A variety of other 
estimators for the same quantity could be tried, but we did not pursue this electrostatic investigation any further at this 
stage.

\begin{table}[ht]
\caption{Conversion table of MC time ($\tau$, MC steps) into real time ($\delta t$, ns). $\Lambda$ is the amplitude of the MC 
step along each Cartesian coordinate. The error bars on computed properties such as the acceptance ratio $\langle \xi \rangle$ 
of MC steps and the diffusion coefficient in MC time are listed explicitly. The error bar on the derived quantity $\delta t / 
\delta \tau$ is indicated implicitly by the number of digits in the quoted result. The conversion from scaled units to real 
units is carried out by assuming $d_{++}=4.25$ \AA\ , and a constant real time diffusion coefficient $D$ of $1 \ 10^{-5}$ 
cm$^2$/s or $100$ \AA\ $^2$/ns for all samples, qualitatively representative of the average diffusion constant of cations and 
anions of light alkali-halide (NaCl, KCl) salts in water.
}
\label{diffuse}
\begin{center}
\begin{tabular}{|c|c|c|c|c|}
    \hline
 M    & $\Lambda/d_{++}$ & $\langle \xi \rangle$ & D$_{MC}$ [d$_{++}^2$/$\tau$] &  $\delta t / \tau$ [ns / MC step] \\
    \hline
 2    &       10         &     $0.20 \pm 0.02$   &   $1.60 \pm 0.05$           &        $    0.28 \pm 0.01$       \\ 
 1    &       10         &     $0.36 \pm 0.04$   &   $2.89 \pm 0.04$           &        $    0.52 \pm 0.02$       \\
 0.5  &       10         &     $0.50 \pm 0.03$   &   $3.97 \pm 0.04$           &        $    0.71 \pm 0.01$       \\
 0.1  &       10         &     $0.72 \pm 0.03$   &   $5.80 \pm 0.05$           &        $    1.04 \pm 0.03$       \\
 0.01 &       10         &     $0.90 \pm 0.04$   &   $7.30 \pm 0.05$           &        $    1.31 \pm 0.02$       \\
\hline
 2    &       19.4       &     $0.18 \pm 0.03$   &   $5.56 \pm 0.08$           &        $     1.005 \pm 0.02$     \\ 
 1    &       14.2       &     $0.34 \pm 0.04$   &   $5.56 \pm 0.06$           &        $     1.004 \pm 0.02$     \\
 0.5  &       11.9       &     $0.49 \pm 0.04$   &   $5.54 \pm 0.08$           &        $     0.996 \pm 0.02$     \\
 0.1  &        9.8       &     $0.72 \pm 0.02$   &   $5.55 \pm 0.05$           &        $     1.003 \pm 0.01$     \\
 0.01 &        8.6       &     $0.90 \pm 0.03$   &   $5.58 \pm 0.07$           &        $     1.008 \pm 0.02$     \\
\hline
\end{tabular}
\end{center}
\end{table}

\begin{figure}[!htb]
\begin{minipage}[c]{\textwidth}
\vskip 0.7truecm
\begin{center}
\includegraphics[scale=0.80,angle=-0]{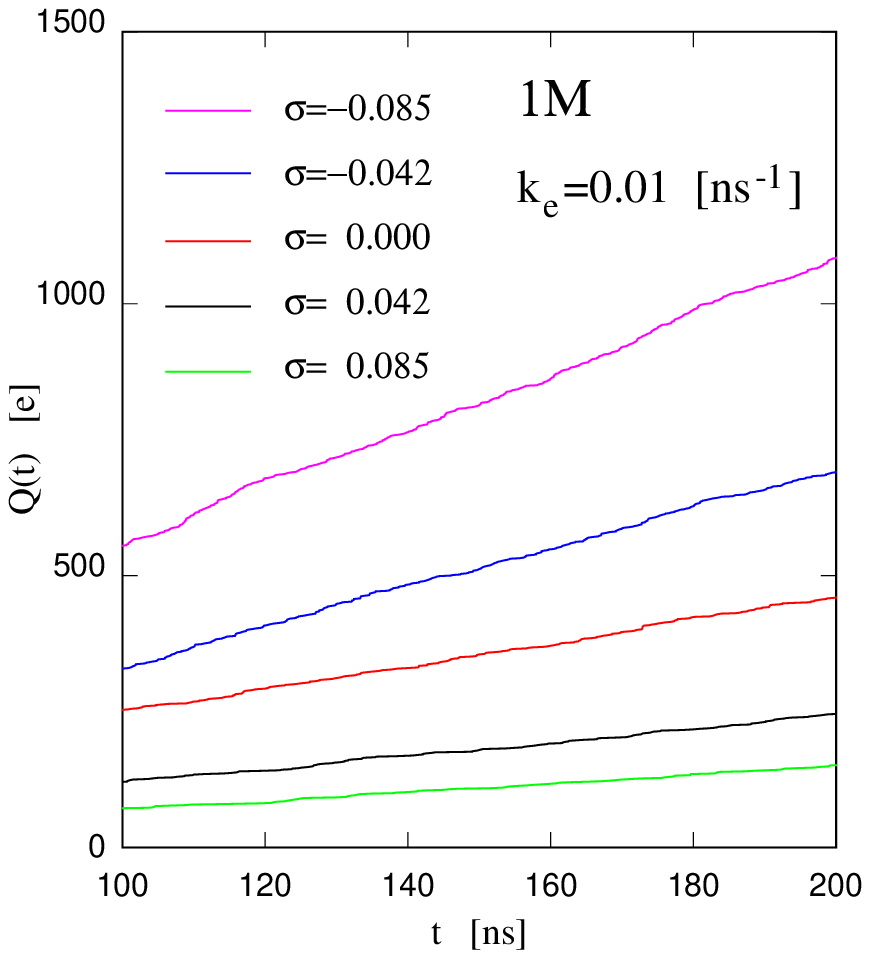}
\vskip 2.0truecm
\caption{Number of electrons transferred from the electrode to the cations as a function of time, across an interface of 
$9460\ d_{++}^2$ area. Simulation carried out on a sample consisting of about 8000 ions at 1M concentration. To show that the 
flow of charge is fluctuating, a short portion (100 ns after the 100 ns equilibration of the flow) of the trajectories has been 
selected, out of runs covering about $2.5$ $\mu$s each. The rate of electron transfer per active ion is $k_e=1\times 10^{-2}$ 
ns$^{-1}$.
}
\label{Qoft}
\end{center}
\end{minipage}
\end{figure}

The electrostatic potential difference across the interfacial double layer may affect the electron transfer rate in a variety of
ways, some of which are covered by the model, and other that are instead excluded at this stage, although they might be added by
model 
refinements.  First of all, $\langle \varphi(R_e)-\varphi(0)\rangle$ is directly related to the charge density profile and, in
particular, to its value at contact, as can be seen in Tab.~\ref{electros}. Moreover, the electrostatic potential contributes to
the electrochemical potential $\bar{\mu}=\mu+\varphi$ for the ions but also for the electrons, thus shifting the energy of 
filled and empty states centred on the ions, and affecting their alinement with the corresponding levels on the electrode side. 
This effect, not included in the present model, could affect the charge transfer especially in the case of transition metal 
electrodes, which, because of narrow d-bands, might show large variations of the electronic density of states over a small 
energy change. The modulation and fluctuations of the potential and electric field close to the interface might have a more 
subtle effect, since they might change the activation barrier for the electron transfer from the electrode to the ions,
significantly amplifying or quenching the rate of the process. These effects can be modelled and their quantitative importance 
can be investigated by simulation. At this validation stage, however, they have been neglected, also because tuning the model
and estimating the electron transfer parameters from the electronic structure requires a more detailed and quantitative picture 
of the system.

\begin{figure}[!htb]
\begin{minipage}[c]{\textwidth}
\vskip 0.7truecm
\begin{center}
\includegraphics[scale=0.80,angle=-0]{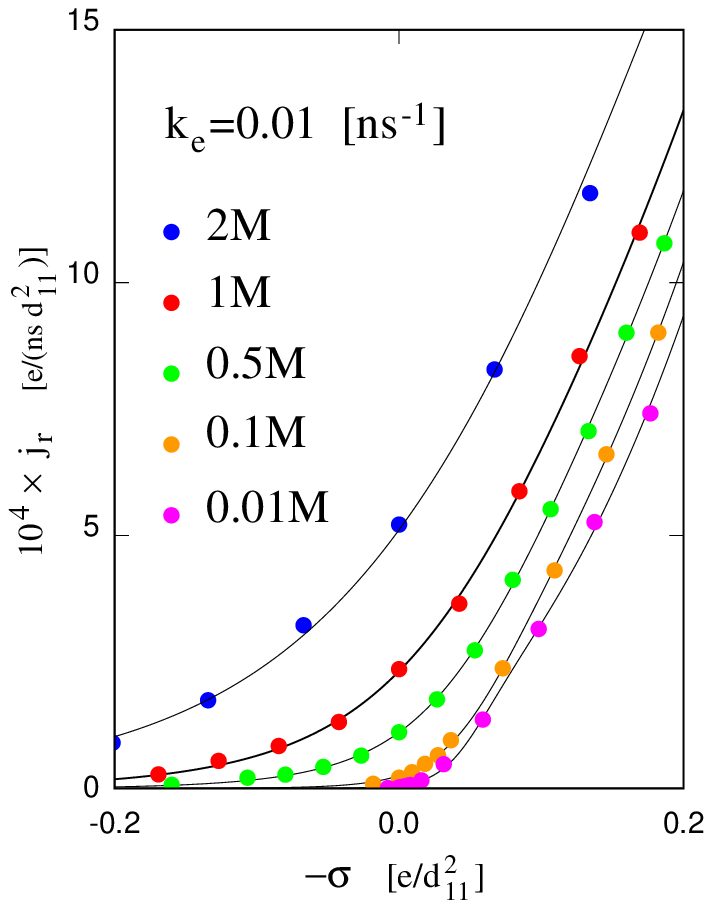}
\vskip 2.0truecm
\caption{Modulus $j_r$ of the radial component of the current density ${\bf j}$ across the metal / electrolyte interface as a 
function of the surface charge density $\sigma$ and molarity M. Dots: simulation results. Full line: interpolations by the 
function in Eq.~\ref{fitjos}.
}
\label{jofS}
\end{center}
\end{minipage}
\end{figure}

A further quantity relevant for the interpretation of the results for the electron transfer simulation is the differential
capacitance of the interface $C_D$, defined as:
\begin{equation}
C_D(\sigma)=\frac{\partial Q}{\partial \phi_d}(\sigma)
\end{equation}
$C_D$ has been determined by fitting the $\sigma$ dependence of the $\langle \varphi(R_e)-\varphi(0)\rangle$ values collected in
Tab.~\ref{electros} and then differentiating the fit. The results for $C_D(\sigma=0)$ are also given in Tab.~\ref{electros}. In 
the present context, the determination of $C_D$ is important because knowing the differential capacitance of the interface 
allows to establish a one-to-one relation between surface charge and potential at the electrode. This, in turn, provides a way 
to translate the results obtained under charge control condition to the voltage control condition that directly corresponds to 
the experiments.

\subsection{Calibration of the diffusion kinetics of ions}
\label{gauge}

To set the stage for the electron transfer simulations, the diffusion of ions in MC time has been computed for samples of the 
same M=0.01; 0.10; 0.5; 1; 2 molar concentrations as in the previous subsections. Ideally, one is looking for the relation 
between the MC and real time diffusion for a given force field model (i.e., the primitive model of electrolyte solutions), 
independently from the the specific application and unaffected by the presence of the interface and by its curvature. For this 
reason, this computation has been carried out on homogeneous samples consisting of 4000 neutral ion pairs in a cubic simulation 
box with periodic boundary conditions applied. To avoid the complication of dealing with the discontinuous change in the number 
and identity of particles implied by the GCMC approach, this part of the simulations has been carried out using the canonical 
ensemble MC, excluding the grand canonical moves. The volume is fixed at the value corresponding to $4000$ ion pairs at the 
target molarity. 

\begin{figure}[!htb]
\begin{minipage}[c]{\textwidth}
\vskip 0.7truecm
\begin{center}
\includegraphics[scale=0.70,angle=-0]{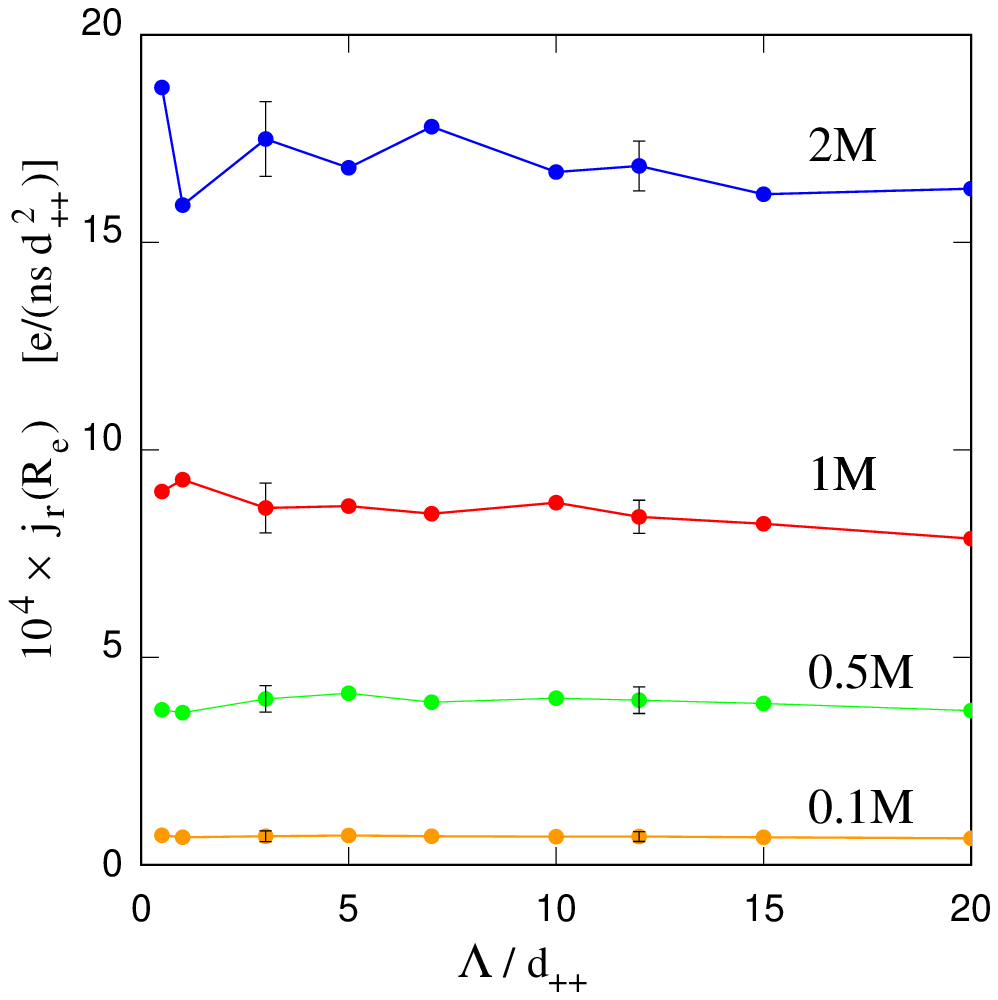}
\vskip 2.0truecm
\caption{Radial current density $j_r(R_e)$ through the electrode/electrolyte solution interface for all the simulated samples
as a function of the amplitude $\Lambda$ of the attempted MC displacement of single ions. For each choice of $\Lambda$ and
molar concentration M, the time scale of the simulation has been gauged according to the protocol described in Sec.~\ref{gauge},
starting from the computation of the diffusion coefficient $D_{MC}$ in MC time. Representative error bars are reported.
}
\label{jrWide}
\end{center}
\end{minipage}
\end{figure}

For the sake of computational convenience, during this preliminary investigation of ionic diffusion, the potential energy has 
been computed using the minimum image convention (mic), which is the standard choice with short range interactions, while long 
range potentials, like the Coulombic one, in extended periodic systems are routinely dealt with using Ewald summation methods. 
The choice of mic, which, incidentally, is the same as the one of Refs.~\onlinecite{torrie, valleau}, is partly justified by the
fact that the side $L$ of each sample is at least ten times longer than the screening length, qualitatively estimated by the 
Debye-H{\"u}ckel 
length $\lambda_{DH}$, reported in Tab.~\ref{activity}. A (slight) exception to this $L/\lambda_{DH} > 10$ rule has to be noted 
only for the most dilute sample (0.01M concentration), for which screening is particularly ineffective. Moreover, diffusion 
properties are more sensitive to 
the short range part of the interparticle potential than to its long range tail, which, instead, might be more important for the
thermodynamic properties of the simulated samples. Remarkably, we observe that, at equal number of particles, i.e., 4000 ion 
pairs, the same value of mean activity coefficient $\gamma_{\pm}$ that gives the target molarity in the neutral and charged 
spherical samples, results in the same molarity for the cubic boxes simulated under the mic boundary condition. This close 
correspondence suggests that even the thermodynamic properties, which are likely to be the most affected by the boundary 
conditions, are not significantly spoiled by the error due to the minimum image convention.

The results for the diffusion coefficient $D_{MC}$ are reported in Tab.~\ref{diffuse} for the samples of molarity $0.01 \leq M 
\leq 2$. As already said, $D_{MC}$ depends significantly on the step size $\Lambda$ for the attempted displacement moves. Two 
values of $\Lambda$ have been considered: i) $\Lambda=10 d_{++}$, which achieves an acceptance ratio of $0.2 \leq \xi \leq 0.6$
for $0.01 \leq M \leq 2$, and, ii) for each molarity, the $\Lambda$ such that $D_{MC}=D$, for which a MC sweep over all 
particles correspond to $1$ ns in real time. These two choices illustrate the fact that one could choose the MC step {\it a 
priori}, using the computed $\delta t/\delta \tau$ to translate MC time in real time units, or, for each concentration of the 
electrolyte, one could figure out what is the MC step that provides a given fixed value of the time conversion ratio. The first 
choice, i.e., a fixed $\Lambda=10 d_{++}$. is the one used for the simulations whose results are presented below. It turns out 
that there is no unique choice for $\Lambda$ and the two sets of values given in Tab.~\ref{diffuse} are only representative of a
whole range of acceptable values. We verified that a broad choice of $\Lambda$ can be combined with the corresponding value of 
$\delta t/\delta \tau$ to give a virtually equivalent description of the flow of electrons through the interface. Quantitative 
support to this statement is given in Sec.~\ref{transfer} below. It is likely that this freedom in selecting $\Lambda$ is 
enjoyed only by models of dilute systems, whose MC sampling of the phase space is weakly dependent on the choice of the MC step.
Denser samples might offer a much narrower choice for the $\Lambda$ and $D_{MC}$, limited by low acceptance ratio for large 
$\Lambda$ or prohibitively slow diffusion in MC time for short $\Lambda$.

\begin{figure}[!htb]
\begin{minipage}[c]{\textwidth}
\vskip 0.7truecm
\begin{center}
\includegraphics[scale=0.70,angle=-0]{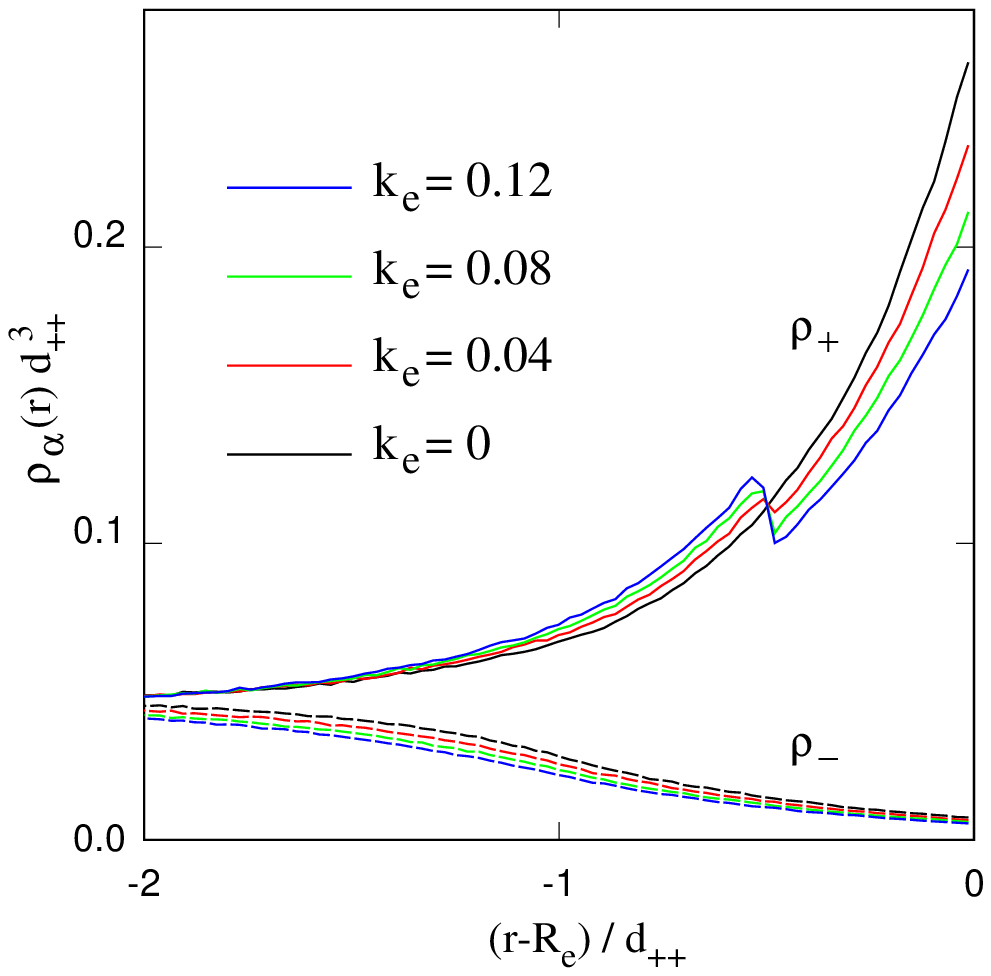}
\vskip 2.0truecm
\caption{Dependence of the density profiles of cations ($\rho_+$: full lines) and anions ($\rho_-$: dash lines) on the rate of 
electron transfer per active atom. The discontinuities in $\rho_+(r)$ mark the boundaries of the active layer at $r=(R_e-
\delta R)$ and $r=R_e$, with $\delta R=d_{++}/2$. The anion density profile $\rho_-$ is continuous with a slight discontinuity
in its derivative which cannot be appreciated on the scale of the figure.
}
\label{rhoflux}
\end{center}
\end{minipage}
\end{figure}

To give an idea of the advantage of this kinetic interpretation of MC enjoys with respect to molecular dynamics, let us consider
the choice of $\Lambda$ such that $D_{MC}=D$ reported in Tab.~\ref{diffuse}. This choice implies that a sweep over particles, 
covers $1$ ns. This is achieved with step amplitudes that provide an acceptance ratio well in excess of $\xi=0.1$, which often 
is considered the minimum MC standard. Moreover, since energies are updated and not recomputed after each attempted move, the 
computational cost of one full sweep over ions is comparable to a single computation of the potential energy from scratch. Then,
taking into account that usual MD time steps are of the order of the fs, it is easy to realise that the efficiency advantage of 
MC over MD is of the order of five to six orders of magnitude. To a large extent, this excellent performance of MC is due to the
dilute state of the samples, that allows MC steps of (nearly) arbitrary length with non-vanishing acceptance probability. 
Although it certainly cannot be generalised, a large advantage of the MC time evolution over the MD one is likely to be enjoyed 
over a broader range of models and simulation conditions than explored in the present study.

\subsection{Electron transfer simulations}
\label{transfer}
The MC simulations of the spherical samples have been extended to include the moves that represent the electron transfer from 
the electrode to the active cations, i.e., those within the thin active layer of width $\delta R=d_{++}/2$ at the interface. Two
series of simulations have been carried out. In one case, attempted displacement and electron transfer moves are accompanied by 
grand-canonical moves that sample different numbers of ions in the system. Attempted displacement and grand canonical moves 
alternate each other at random, with the latter being tried, on average, every $10^3$ attempted displacement ones. 
Simulations belonging to the second sequence exclude the purely grand canonical moves, and include only attempted displacements 
and electron transfer events. In all cases, the electron transfer occurs at a time selected from the exponential probability 
distribution as explained in Sec.~\ref{method}. Including grand canonical moves, even at low rate, has the advantage that the 
simulated system is in equilibrium with a known bulk state, defined by its excess chemical potential. Excluding these moves, 
instead, has the advantage of making more precise the relation between diffusion and time that underlies the approach being 
proposed. It turns out that all aspects of the simulated systems provided by the two algorithms are indistinguishable from each 
other. In what follows all data and plots are from the simulations without the grand canonical moves.

\begin{figure}[!htb]
\begin{minipage}[c]{\textwidth}
\vskip 0.7truecm
\begin{center}
\includegraphics[scale=0.70,angle=-0]{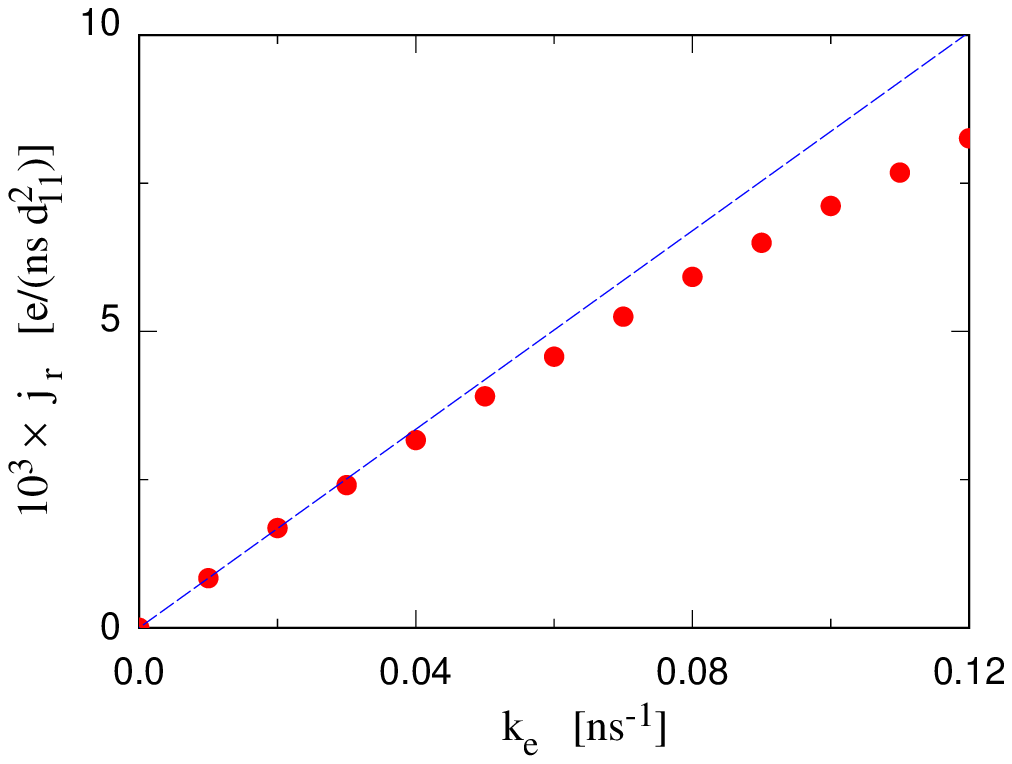}
\vskip 2.0truecm
\caption{Dependence of the radial current density $j_r(R_e|k_e)$ through the interface as a function of the kinetic coefficient 
$k_e$. The straight line is the tangent to $j(R_e | k_e)$  at the origin, and has been drawn to highlight the non-linearity of 
the simulation data.
}
\label{jofk}
\end{center}
\end{minipage}
\end{figure}

Following the protocol outlined in Sec.~\ref{method}, simulations start from a configuration well equilibrated by the GCMC
method. All active ions are identified and given an expiration time, whose common origin is the system time $t_s=0$. After this
initial time, the synchronisation of the ion lifetimes is quickly lost, with some active ions being neutralised, and new
ones joining the active class at virtually random times. The beginning of the simulation, when the aging of active ions is
synchronised and the density profile of the ions at the interface has not been affected yet by the flow of charge, might differ 
from the stationary state of charge transfer and diffusion of ions through the system, when ions become active (by their random
translation) and undergo neutralisation at uncorrelated times. To prevent biasing the results, a transient stage of about 100 ns 
is discarded from the time dependent simulation before starting to accumulate statistics. The transfer of electrons to the 
electrolyte cations, which is the subject of the investigation, is enhanced when the fluid side is positively charged and the 
surface charge is negative. For this reason, a few figures discussed in this section and in SI report properties as a function 
of $-\sigma$ instead of $\sigma$.

The first quantity that we plot is the integral over time of the charge transferred from the electrode to the cations, that is
shown in Fig.~\ref{Qoft} as a function of time for different values of the surface charge density. The electrolyte concentration
in the sample is 1M, and the rate of electron transfer per active ion is fixed in all cases at $k_e=1\times 10^{-2}$ ns$^{-1}$.
Similar simulations have also been performed for all the other samples considered in this study, whose molarity goes from 0.01M
to 2M. 

The overall electron transfer rate per unit area at the interface, which represents a radial current density 
$\langle j_r(R_e)\rangle$, is obtained as the time derivative of the electron charge $Q(t)$ transferred per unit area. On 
average, the tangential components $(j_{\theta}; j_{\phi})$ of $\langle {\bf j}\rangle$ vanish by symmetry. The simulation 
results collected in Fig.~\ref{jofS} show the dependence of $\langle j_r\rangle$ on the surface charge $\sigma$ for all the 
samples of different electrolyte concentration. The strong 
dependence of $\langle j_r\rangle$ on negative values of $\sigma$ reflects primarily the similar dependence on $\sigma$ of the 
average surface density of active ions $\langle n_A \rangle$. This close relationship is highlighted by Fig.~7 in Sec.~S5 of
SI, plotting $\langle n_A \rangle$ as a function of $\sigma$ for different molarities of the electrolyte, whose similarity with
Fig.~\ref{jofS} is apparent. In turn, the dependence of $\langle j_r \rangle$ on $\langle n_A \rangle$, which, according to 
Eq.~\ref{contc} grows quadratically with increasing $-\sigma$ values, suggests to fit both quantities with the functional form:
\begin{equation}
f(\sigma)=A (1+\sigma^2)\frac{\exp{(B\sigma)}}{1+C\exp{(B\sigma)}}
\label{fitjos}
\end{equation}
where $f(\sigma)$ is either $\langle j_r \rangle$ or $\langle n_A \rangle$, and $A$, $B$ and $C$ are the fit variables. This 
expression joins the quadratic grows of $f(\sigma)$ at large $-\sigma$ with a slow exponential-type decrease at large positive
$\sigma$, when the electrostatic repulsion depletes the active layer of reactive ions. The fit is remarkably good, and, it has 
been verified that the overall scale parameter $A$ depends very sensitively on the molarity of the solution and on the electron 
transfer rate $k_e$. The parameters $B$ and $C$ reflect primarily the screening properties of the electrolyte close to the 
interface. Notice that $\sigma=0$, corresponding to the potential of zero charge  (PZC) for the electrode, does not imply zero 
current, whose flow is dictated by the electron energy diagram in Fig.~\ref{scheme}. 

\begin{figure}[!htb]
\begin{minipage}[c]{\textwidth}
\vskip 0.7truecm
\begin{center}
\includegraphics[scale=0.70,angle=-0]{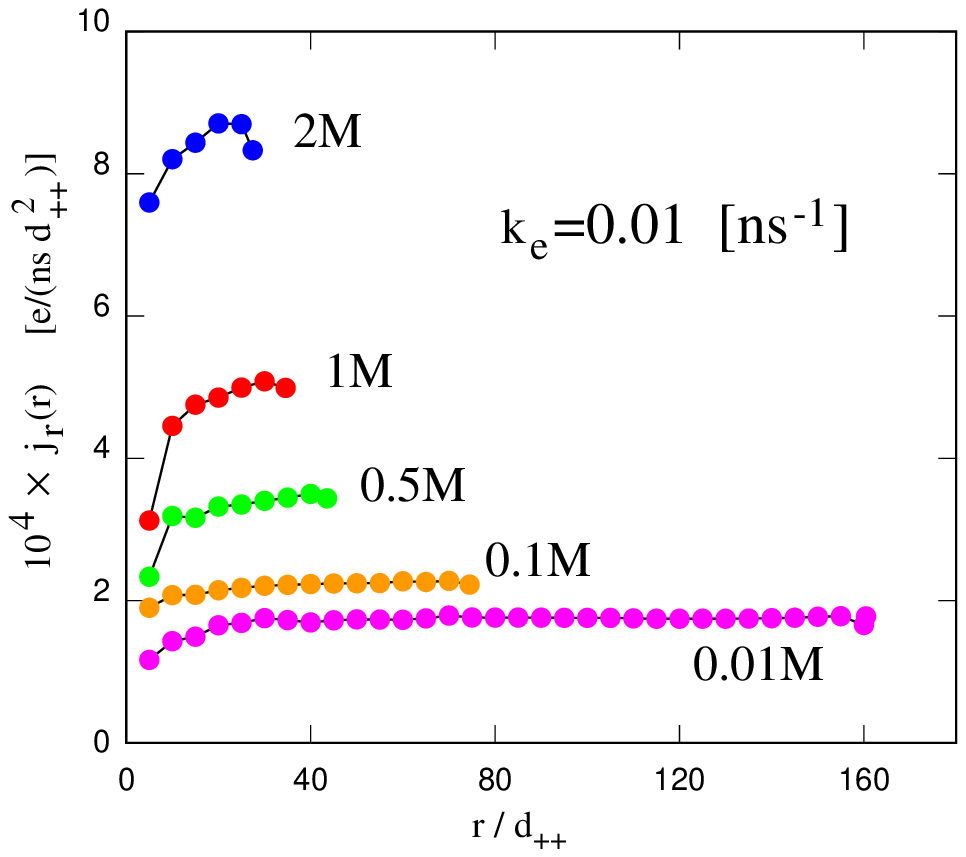}
\vskip 2.0truecm
\caption{Dependence of the radial component $j_r(r)$ of the current on the distance  $r$ from the centre of the spherical 
electrode (dots). The black full line is a guide to the eye. The last point of each curve corresponds to the 
electrode/electrolyte solution interface. While at the interface the charge flow is due to electrons neutralising cations,
in the rest of the sample there is an ionic current of cations drifting towards the cathode.
}
\label{constantj}
\end{center}
\end{minipage}
\end{figure}

The $\langle j_r (\sigma) \rangle$ curves for the different electrolyte concentrations shown in Fig.~\ref{jofS} trivially 
converge to the zero current condition for large $\sigma$ values, because of the vanishing of $\langle n_A \rangle$ in the same 
limit. More interestingly, the $\langle j_r (\sigma) \rangle$ curves seem to converge to a unique limiting curve also in the 
opposite limit of large 
$-\sigma$ values, in which the average surface density of active ions $\langle n_A \rangle$ becomes uniquely determined by 
$\sigma^2$, as apparent from the relation in Eq.~\ref{contc}, at the same time as its dependence on the bulk osmotic pressure 
becomes negligible.

It is important to emphasise that the precise choice of the MC step $\Lambda$ is not crucial to provide the correct description 
of the electron transfer kinetics, at least over a broad interval of values, and {\it provided} the conversion of MC time to
real time is carried out according to the approach described in Sec.~\ref{method} and \ref{gauge}. To verify this important 
point, simulations have been carried out for all electrolyte concentrations already considered, varying the MC step $\Lambda$
and converting time with the $\Lambda$ and concentration dependent factor $\delta t/\tau$ that has been computed using the 
protocol specified in Sec.~\ref{gauge}. 
The results, shown in Fig.~\ref{jrWide}, confirm that the choice of $\Lambda$ is not unique, and produces equivalent results 
provided the set of (Molarity, $\Lambda$ and $\delta t/\tau$) values are consistent. This feature allows to select $\Lambda$ in 
such a way to optimally balance the efficiency (requesting wide $\Lambda$ and long $\delta t/\tau$) as well as the time 
resolution
(requesting narrow $\Lambda$ and short $\delta t/\tau$) of the MC evolution of the system. As already stated, there is no 
guarantee that this broad freedom in choosing $\Lambda$ is shared by more realistic models of electrode/electrolyte interfaces, 
but even a narrower choice is likely to be compatible with the simulation study of a sizeable class of systems and models.

\begin{figure}[!htb]
\begin{minipage}[c]{\textwidth}
\vskip 0.7truecm
\begin{center}
\includegraphics[scale=0.80,angle=-0]{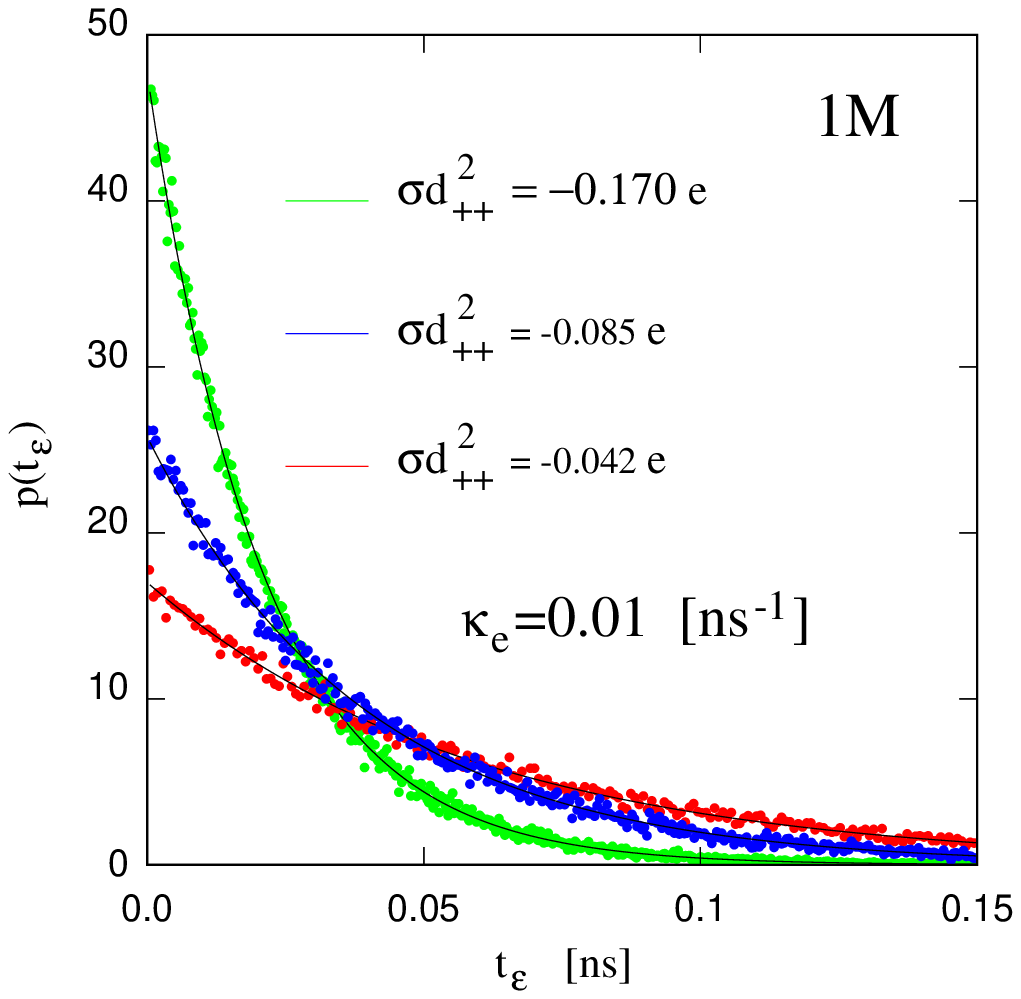}
\vskip 2.0truecm
\caption{Probability distribution for the time separation $t_{\epsilon}$ between successive electron transfer events. Sample of 
(nominal) 8000 ions at 1M concentration. The small dots in colours represent the simulation results. The full lines (black) 
correspond to the analytical result expressed by Eq.~\ref{tepsi} with $\langle n_A \rangle$ computed by simulation for the three
samples of different surface charge density $\sigma$.
}
\label{histo}
\end{center}
\end{minipage}
\end{figure}

As repeatedly pointed out, the dependence of $\langle {\bf j}(R_e) \rangle \equiv j_r(R_e)$ on $\sigma$ is due primarily to the 
dependence of the contact density, and therefore of the number of active ions, on $\sigma$. An additional dependence might
come from the fact that 
the number of active ions depends on the current density $\langle {\bf j}\rangle$ itself, which tends to deplete the ion density at contact with the electrode. This negative feedback mechanism is unambiguously confirmed by the simulation results.
To emphasise this effect, charge transfer simulations have been carried out for systems whose surface charge density $-\sigma$
is large, exploring a range of $k_e$ also extending up to fairly large values. The results are shown in Fig.~\ref{rhoflux}. It 
is apparent that with increasing rate $k_e$, the density of cations in the active layer is increasingly depleted, while the 
density of anions in the same range is nearly unchanged. Just outside the active layer, i.e., for $r \leq R_e-\delta R$, the 
cation density is slightly enhanced, since the screening of the interfacial charge is less effective when the electron current 
is flowing. As a consequence of this depletion, the dependence of overall transfer rate ${\bf j}$ on the $k_e$ rate constant is 
slightly sub-linear, as shown in Fig.~\ref{jofk}.

\begin{figure}[!htb]
\begin{minipage}[c]{\textwidth}
\vskip 0.7truecm
\begin{center}
\includegraphics[scale=0.70,angle=-0]{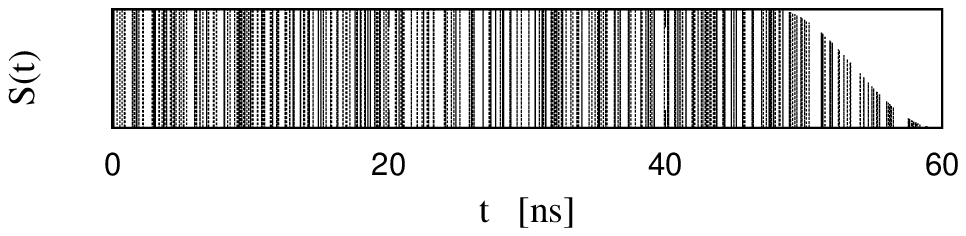}
\vskip 2.0truecm
\caption{Time series giving the exact time of each electron transfer in the sample of Fig.~\ref{Qoft}. The time series has been
weighted by a windowing function equal to 1 up to 80 \% of the total interval, and smoothly going to zero in the last 20
For the sake of simplicity, the figure reports a time series much shorter than the one used to compute the power spectrum.
}
\label{tser}
\end{center}
\end{minipage}
\end{figure}

The choice of re-inserting the charge of the neutralised ions at a random position in the system volume according to the 
probability distribution $p_r(r)$ in Eq.~\ref{constj} aims at obtaining a radial component $j_r({\bf r})$ of the current that is
independent from position and matches the flow of charge at the electrode/electrolyte solution interface. This requirement
has been verified by considering a sequence of spherical surfaces concentric with the sample whose radii are spaced by a 
regular distance of $5 d_{++}$. Then, the number and direction of cations and anions crossing each surface during the simulation
is recorded, and transformed into the radial current $j_r(r)$ which represents the net transfer of ions across each unit surface
($d_{++}^2$) during a unit of (real) time. The results of this analysis of trajectories is reported in Fig.~\ref{constantj}. It 
is apparent that the requirement of uniform $j_r$ across the system is satisfied fairly well. A non-negligible deviation only 
occurs close to the origin, due to the transformation of $p_r(r)=1/r$ into $p(r)=1/(1+r)$ introduced to avoid dealing with a 
distribution which is singular at the origin.

As already pointed out, achieving a (nearly) uniform radial current ($j_r({\bf r})=k$) implies a violation of the continuity 
equation. After a short transient, in fact, $\partial \rho_Q({\bf r}) /\partial t=0$ everywhere, while $\nabla \cdot 
{\bf J}({\bf r})=2k/r$, thus:
\begin{equation}
\nabla \cdot {\bf J}({\bf r}) +\frac{\partial \rho_Q({\bf r})}{ \partial t}=2k/r \neq 0 
\end{equation}
In these expressions, $\rho_Q({\bf r})=e\left[ \rho_+({\bf r})-\rho_-(({\bf r})\right]$, where $e$ is the electron charge. The 
violation of the continuity equation, however, cannot be avoided for this kind of processes, in which the interface is a sink of
(in this case) cations. Since the system evolves under stationary conditions, the sink of cations needs to be compensated by an 
equivalent source, that also violates the continuity equation, and that in the model is represented by the re-introduction of 
the cations neutralised at the interface. Then, requesting $j_r(r)=k$ is only a way to decide where and how strongly the 
violation affects the system. The other extreme choice would be to enforce the continuity equation {\it almost everywhere}, that
however requires $j_r({\bf r})=k'/r^2$. In turn, this implies that the violation of the continuity equation is concentrated at 
the interface $r=R_e$ and at the origin $r=0$. In this case, in addition, $j_r({\bf r})$ is singular at the origin. Considering 
these two cases, we think that the $j_r({\bf r})=k$ choice is the least problematic one. In this respect, the planar case is 
somewhat easier, because in such a case the violation of the continuity equation can be confined at the two opposite interfaces.

Similarly to what has been found for structural and electrostatic properties, at equal charge density $\sigma$, the dependence 
of the current density on the size of the spherical sample is minor. Quantitative data are given in Sec.~S6 of SI.

In the model of electron transfer described above, the rate constant $k_e$ accounts for all contributions to the activation
barrier that opposes the electron transfer itself. More refined models could be built by making the description of the barrier 
and of its different components more detailed. For instance, one could isolate the reorganisation energy from the other 
contributions, introducing a step-like potential that opposes the penetration of cations and anions into the active range.
This simple addition would already give a more complex and richer model to simulate.

A crucial aspect of the proposed approach is that, at variance from the deterministic, continuous and delocalised picture
of DFT-based approaches whose Kohn-Sham orbitals do not strictly correspond to real electrons, the model retains a discrete
representation of both the electrons and their transfer across the interface. In particular, each electron transfer is a
discrete event taking place at time $t_i$,, whose sequential timing represents a time series $\{ t_i, i=1, ..., \Pi \}$, whose 
statistics can be analysed in much detail.

\begin{figure}[!htb]
\begin{minipage}[c]{\textwidth}
\vskip 0.7truecm
\begin{center}
\includegraphics[scale=0.70,angle=-0]{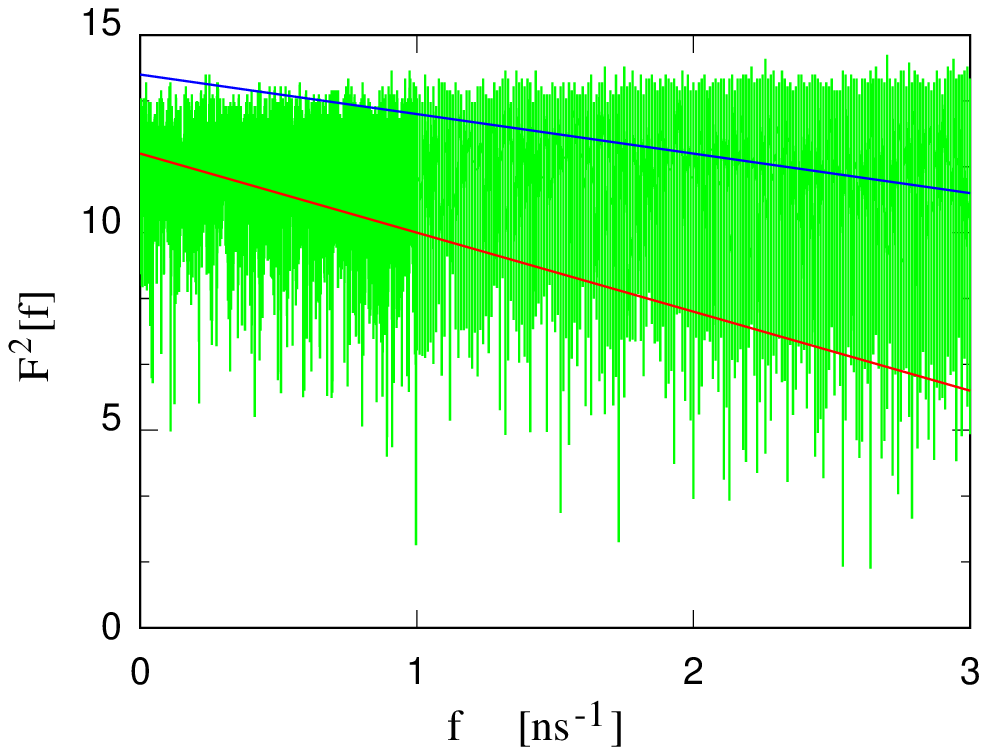}
\vskip 2.0truecm
\caption{Power spectrum (green line) of the time series giving the exact time of each electron transfer in the sample of 
Fig.~\ref{Qoft}. The power spectrum is computed as the square of the Fourier transform of the time series. Red line: line of
slope $m=-2$. Blue line: line of slope $m=-1$.
}
\label{flick}
\end{center}
\end{minipage}
\end{figure}

The simplest analysis concerns the probability distribution of the separation $t_{\epsilon}$ between consecutive electron 
transfer events, shown in Fig.~\ref{histo}. Assuming, at first, that the time series is generated by a single active site, which
remains unaltered by the transfer and continue to be active, the separation between successive events is precisely the random 
variable $\Delta t$ whose distribution is given by Eq.~\ref{decay}. The probability distribution for the time separation of
transfers taking place over the whole interface, however, is less predictable, since it results from the superposition of many 
exponential time series (as many as the number $n_A$ of active ions), whose origin is not synchronised, and whose number is
fluctuating in time. Nevertheless, the result conforms to the simplest possible guess: the average time separation is
$\langle t_{\epsilon} \rangle= 1/( \langle n_A\rangle k_e)$ and the probability distribution is:
\begin{equation}
p(t_{\epsilon})= \bar{K}_e \exp{(-\bar{K}_e t_{\epsilon})}
\label{tepsi}
\end{equation}
with $\bar{K}=\langle n_A\rangle k_e$. This result is illustrated in Fig.~\ref{histo}.

A more refined statistical analysis could give insight into the relaxation processes that take place in the system, and also
allow a connection with experimental observations. To carry out this analysis, the time series is written as:
\begin{equation}
{\cal{J}}(t)=\sum_{i=1}^{\Pi} y(t)\delta(t-t_i)
\end{equation}
that represents the instantaneous current across the whole interface at time $t$. The function $y(t)$ is a suitable windowing 
function introduced to make Fourier-representable a sequence that otherwise is non-periodic. Its definition and application are
illustrated in Fig.~\ref{tser}. The power spectrum, computed as the square of the Fourier transform of ${\cal{J}}(t)$, is
shown on logarithmic scales in Fig.~\ref{flick}. Notice that the time series transformed to obtain the power spectrum is much 
longer than the one given in Fig.~\ref{tser}, being $30$ $\mu$s long. The result is somewhat ambiguous, since Fig.~\ref{flick}
shows both a constant {\it white noise} background and a component which behaves like $1/f^{\alpha}$, with $1 \leq \alpha 
\leq 2$, the power spectrum contains aspects reminiscent of {\it flicker noise}. These observations are highlighted by the two
straight lines of slope $-1$ and $-2$ superimposed to the noise power spectrum in Fig.~\ref{flick}.

\begin{figure}[!htb]
\begin{minipage}[c]{\textwidth}
\vskip 0.7truecm
\begin{center}
\includegraphics[scale=0.80,angle=-0]{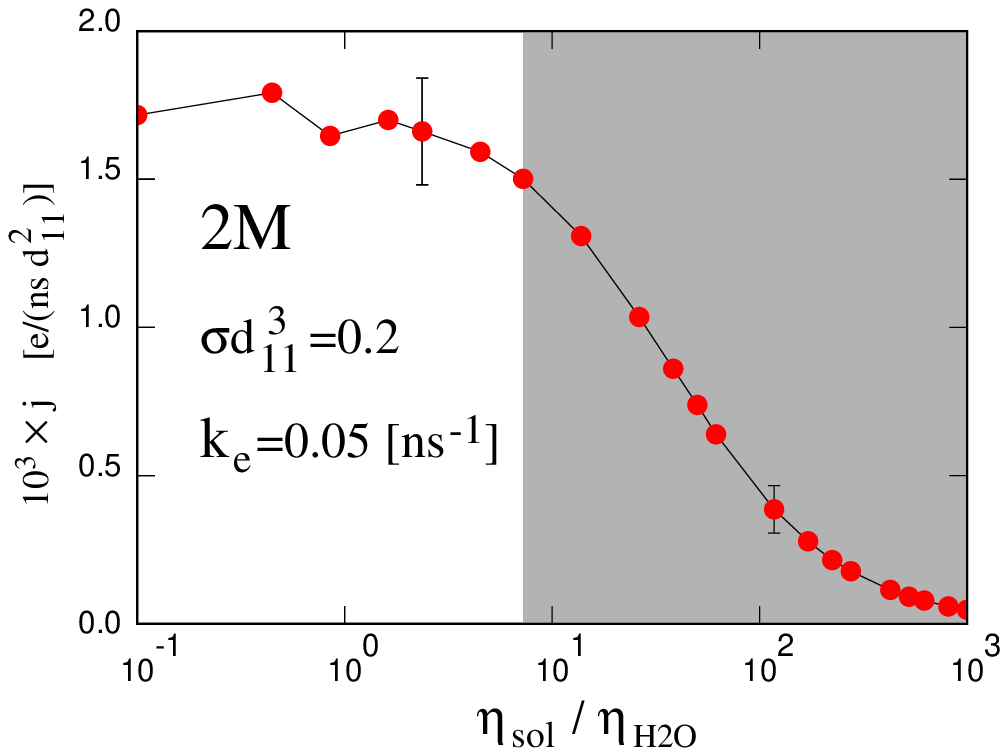}
\vskip 2.0truecm
\caption{Radial component $j_r$ of the current density across the electrode / electrolyte solution interface as a function of 
the viscosity $\eta_{sol}$ of the solvent in which the electrolyte ions diffuse. Representative error bars are given. The 
viscosity is measured in units of the water viscosity coefficient $\eta_{H2O}$. The calibration of the real time $t$ to MC time 
$\tau$ conversion relies on the Stokes-Einstein relation to estimate the real-time diffusion coefficient of the ions in the 
solvent of any given viscosity, starting from the $D=10^{-5}$ cm$^2$/s value adopted for their diffusion in water at standard 
conditions, irrespective of concentration in the $0.01 \leq M \leq 2$) range. Dots: simulation results. The full line is a guide
to the eye. The gray area identifies the $\eta_{sol}/ \eta_{H2O} \geq 10$ range in which the electron transfer is limited by 
diffusion.  
}
\label{jofeta}
\end{center}
\end{minipage}
\end{figure}

The second qualifying aspect of our model is its explicit coupling of the two processes that dominate the charge transfer at the
electrode / electrolyte interface. To illustrate this dependence, the calibration of the correspondence between MC and real time
has been repeated assuming that the ions diffuse in a solvent whose viscosity $\eta_{sol}$ varies from $10$\% of the water value
$\eta_{H2O}$ to $2000 \times \eta_{H2O}$. This seemingly wide range is in fact rather moderate. considering that the viscosity 
of honey, for instance, is already $\sim 40\times 10^3\ \eta_{H2O}$. The calibration has been carried out by scaling the 
diffusion constant of the ions using the Stokes-Einstein relation ($D\times \eta = constant$), which, although not 
quantitatively accurate, reproduces the correct trend and order of magnitude for the relatively dilute solutions of the present 
study. Then, simulations have been carried out for solutions of 2M concentrations, with the surface charge density $\sigma=0.2$ 
e/d$_{++}^2$, and electron transfer rate $k_e=0.05$ $ns^{-1}$. In each case the MC step $\Lambda$ for the attempted displacement
of the ions was fixed at $10 d_{++}$, and the time conversion factor was selected as specified in the previous sentences of this
paragraph. The results of these simulations are reported in Fig.~\ref{jofeta}. It is apparent that increasing $\eta_{sol}$, thus
decreasing the diffusion rate in real time, enhances the depletion of the active layer of reactive ions, and decreases the 
overall electron transfer rate. For $\eta_{sol} > 10 \eta_{H2O}$ (corresponding to the gray area in Fig.~\ref{jofeta}), the 
electron transfer rate is diffusion limited, while below this value of $\eta_{sol}$ the process is limited by the kinetic 
parameter $k_e$. The transition between the two regimes is relatively sharp. 

\section{Summary and conclusions}

Simulating the electron transfer at electrochemical interfaces is a crucial goal which could greatly impact multiple conceptual 
and applied research topics. The process encompasses a number of electronic structure aspects, concerning the energy spectrum 
and spatial localisation of filled and empty states on the two sides of the interface, as well as statistical mechanics aspects,
concerning the distribution of ions in the diffuse layer at the interface and the strength and fluctuations of the electrostatic
potential (and corresponding electric field) on the electrolyte solution side. Kinetics is also very important, both for the 
electron transfer process, controlled by activation barriers and affected by quantum and thermal fluctuations, and also for the 
diffusion of ions in solution, which is required for reaching a stationary state during the operation of the electrochemical 
device to which the electrode / electrolyte interface belongs. The electron transfer involves only the active ions in the 
immediate vicinity (order of the \AA\ ) of the electrode, while diffusion involves all the electrolyte ions, whose precise state
can be fixed by applying the GCMC algorithm, which provides the required connection with an extended reservoir of known 
thermodynamic conditions.

Seen from the time scale of diffusive and relaxation processes in the electrolyte solution, the electron transfer is a sudden 
event taking place after a long lag time whose random duration reflects the probabilistic nature of quantum mechanical events.
These aspects orient the choice of the overall simulation method to Monte Carlo, able to deal with discontinuous and
stochastic aspects, and also to bypass the high frequency vibrational aspects that limit the time span of atomistic molecular 
dynamics simulations, retaining diffusion and the relaxation processes that drive the long term evolution of the system. 
Diffusion, in particular, couples the stochastic electron jumps and the atomistic processes that involve the electrolyte ions. 
After a relatively short transient, the mean square displacement of the ions depends linearly on time in real-life diffusion. 
In a similar way, the mean square displacement of the ions depends linearly on the number of steps in MC, which represents the 
closest analogue to an intrinsic time scale in MC simulations. In the present approach, the rate of the two processes are 
connected by imposing that real and MC time are equivalent when they result in the same increase of mean square displacements. 
This gives a way to express the MC evolution in real time units and to synchronise diffusion and electron transfer in the model. 
The model and the simulation protocol introduced in Sec.~\ref{method} reflect all these aspects, although, to highlight the most
important features unencumbered by details, it does it in the simplest possible way. Arguably one of the most crude 
simplification is the fact that the electrolytic solution is described at the implicit solvent level. Together with the moderate
electrolyte concentration (up to 2M), the implicit solvent assumption allows to reposition ions from the interfacial active 
layer to well inside the spherical sample, as required to simulate the kinetics of cations' neutralisation by the transferred 
electrons while: (i) conserving the total sample charge; (ii) sustaining the same radial current density throughout the system, 
from an inner spherical core up to the interface. 

Monte Carlo simulations of the coupled electron transfer and ionic diffusion carried out by the model of Sec.~\ref{method} show 
that, after a relatively short transient, the samples reach a stationary state in which the depletion of active ions is 
compensated by diffusion. The interplay, and, in some instances, the competition between these two mechanisms determines a range
of features concerning the steady state operation of the electrochemical interface. For instance, depending on the electrolyte
solution viscosity, the process presents two regimes (see Fig.~\ref{jofeta}), the first one being diffusion limited, the second 
one being limited by the inherent electron transfer rate per active ion $k_e$. Of course, the diffusion limited regime 
corresponds to low values of the ions' diffusion constant, the second one corresponds to high values of $D$ and $D_{MC}$, and 
the transition between the two regimes is fairly sharp. Moreover, the partial replenishment of the depleted active ions results 
in the sub-linear dependence of the overall transfer rate on the kinetic coefficient $k_e$. 

The implemented model is particularly simple, since it corresponds to the scheme of Fig.~\ref{scheme}, and
neglects the feedback on the electronic levels due to the value and the fluctuations of the electrostatic potential driven by
the interfacial charge and electron flow. In this simplified picture, the primary reason for the dependence of the overall 
transition rate on the electrode surface charge density $\sigma$ is represented by the similar dependence of the number of 
active ions on the surface charge. The resulting $j_r(\sigma)$ function is modelled by the fit in Eq.~\ref{fitjos}, and, at 
large $-\sigma$ values, primarily reflects the contact density theorem expressed by Eq.~\ref{contc}. 

The correspondence between 
the stochastic evolution of the model and real time provides an opportunity to investigate the size and nature of the noise in 
the current flow. The model, however, is likely to be still too sketchy to provide a reliable account of this complex property. Results from the present simulations, plotted in Fig.~\ref{flick}, give an uncertain answer, showing aspects that could suggest 
both a white or a flicker noise signature. Better models, including feedback of the electrostatic potential strength and 
fluctuations on filled and empty electronic states on the ions might change the picture and provide a more definite result.

Besides adding the effect of electrostatics on the (empty and filled) electron energies of ions, many other opportunities exist 
to improve the model, covering a wider range of phenomena and making the approach more predictive. For instance, the further
role of the ion whose oxidation state is being changed at the interface is left unspecified by the present model. However, the
model could be extended by attributing realistic interactions to the transformed species, opening the way to a more 
comprehensive description of electrochemical devices, up to the complexity of Galvanic and fuel cells. Moreover, the model 
neglects the effect of the electric field at the interface on the reaction barrier that determines $k_e$. This effect, however, 
could be modelled and reintroduced, opening the way to verify by simulation the validity and microscopic origin of Volmer-Butler
equation.\cite{bockris} A first step in modeling $k_e$ and its dependence on various contributions could consist in isolating 
the effect of the reorganisation energy from other, more genuinely electronic factors. The reorganisation energy, in 
particular, could be mimicked by a repulsive step potential that could limit the access of ions to the active layer. 

The description of the electronic part of the process could be improved by adding an atomic-like quantum Hamiltonian to the 
active ions, with an explicit coupling with the electron states spilling out of the electrode. This extension is certainly 
ambitious and challenging, but it would open the way to semi-quantitatively including non-adiabatic effects into the electron 
kinetics at the interface.\cite{mohr}

One of the aspects that most urgently requires improvements is the need to carry out random, long distance displacements of ions
from the active layer to well inside the electrolyte solution side of the interface, as described at the end of 
Sec.~\ref{method}. In practice, this can be carried out only at low fluid particles density, but it would hardly be feasible 
with a dense, explicit solvent model. Removing this limitation is challenging, but recent developments (involving the adaptive 
resolution algorithm\cite{robin1, robin2, robin3, robin4}) might indicate the way forward. In 
this method, the portion of the system described in atomistic detail is coupled with an ideal gas reservoir of known chemical 
potential, in which the insertion, removal or long range random displacement of particles can take place with virtually unit 
probability. These features could be slightly adapted to produce a gradient of electrochemical potential, which, in turn, would 
force a stationary flow of current similar to what has been obtained in the present case.

Even before these improvements and extensions will be made, the present scheme and its implementation represent a rich
statistical mechanics model to investigate open systems operating at steady state conditions, mixing quantum and classical 
aspects in their time evolution, whose detailed analysis could shed light on many complex aspects of electrochemical interfaces.

ACKNOWLEDGMENTS: D.V.-D., F. S. and  N.C.F.-M. acknowledge the Deutsche Forschungsgemeinschaft (DFG, German Research Foundation) for funding through the research group FOR 2982–UNODE, Project number 413163866. N.C.F.-M. gratefully acknowledges the ECHELON Project from the Carl Zeiss Foundation for financial support. R.C.-H. acknowledges the European Union for funding through the Twinning project FORGREENSOFT (grant no. 101078989 under HORIZON-WIDERA-2021-ACCESS-03). R.C.-H. and F.S. acknowledge funding from SFB-TRR146 of the German Research Foundation (DFG)–Project No. 233630050. One of us (PB) thanks Andrea Grisafi for useful 
discussions.

\end{document}